%% file: zzz_main_arxiv.tex
\PassOptionsToPackage{table}{xcolor}
\documentclass[acmtog, nonacm]{acmart}
\acmSubmissionID{1318}

\usepackage{wrapfig}
\usepackage{algorithm}
\usepackage{algpseudocode}
\usepackage[makeroom]{cancel}
\usepackage{tabularx}
\usepackage{booktabs}
\usepackage{lipsum}
\usepackage{amsthm}
\usepackage{pifont}

\input{preamble.tex}

\begin{document}

\title{Mesh Processing Non-Meshes via Neural Displacement Fields}

\author{Yuta Noma}
\email{yutanoma@dgp.toronto.edu}
\orcid{1234-5678-9012}
\affiliation{
  \institution{University of Toronto}
  \city{Toronto}
  \country{Canada}
}

\author{Zhecheng Wang}
\email{zhecheng@cs.toronto.edu}
\orcid{1234-5678-9012}
\affiliation{%
  \institution{University of Toronto}
  \city{Toronto}
  \country{Canada}
}

\author{Chenxi Liu}
\email{chenxil@cs.toronto.edu}
\orcid{1234-5678-9012}
\affiliation{%
  \institution{University of Toronto}
  \city{Toronto}
  \country{Canada}
}

\author{Karan Singh}
\email{}
\orcid{1234-5678-9012}
\affiliation{%
  \institution{University of Toronto}
  \city{Toronto}
  \country{Canada}
}

\author{Alec Jacobson}
\email{jacobson@cs.toronto.edu}
\orcid{1234-5678-9012}
\affiliation{%
  \institution{University of Toronto and Adobe Research}
  \city{Toronto}
  \country{Canada}
}

\renewcommand{\shortauthors}{Noma et al.}

\begin{abstract}

\input{sections/00_abstract}

\end{abstract}

\begin{teaserfigure}
    \centering
  \includegraphics[width=\textwidth]{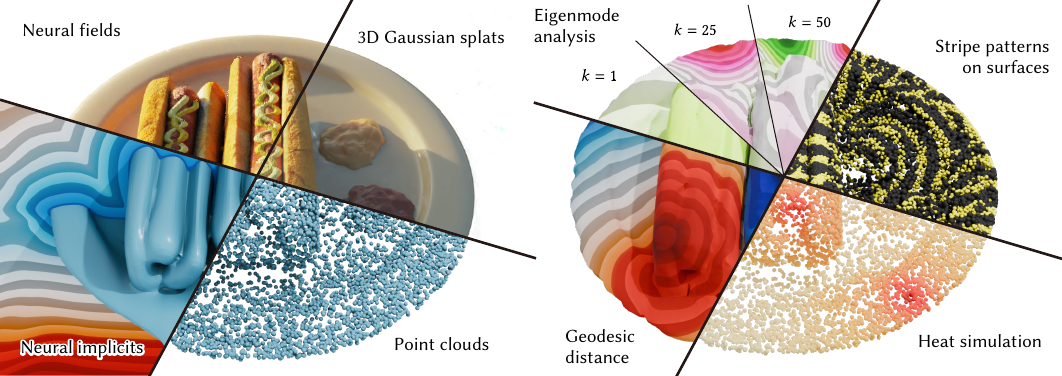}
  \caption{We propose a compact neural map that allows fast and accurate computation of geometry processing tasks on non-mesh surfaces. Our method supports various surface representations, including neural implicits, point clouds, and 3D Gaussian splats. }
  \label{fig:teaser}
\end{teaserfigure}

\maketitle

\input{sections/01_introduction}

\input{sections/02_related-work}

\input{sections/03_method}

\input{sections/04_shape-representations}

\input{sections/20_experiments}

\begingroup
\let\u\relax 
\input{sections/80_conclusion}

\bibliographystyle{ACM-Reference-Format}
\bibliography{references}

\newpage

\input{sections/89_figureonly}

\appendix

\clearpage

\input{sections/90_appendix}

\endgroup

\end{document}

%% file: preamble.tex
\usepackage[mathletters]{ucs}

\usepackage{bm}

\providecommand{\J}{}

\providecommand{\R}{}

\providecommand{\d}{}

\providecommand{\n}{}

\providecommand{\p}{}
\providecommand{\q}{}

\providecommand{\t}{}
\providecommand{\u}{}
\providecommand{\v}{}

\providecommand{\x}{}
\providecommand{\y}{}
\providecommand{\z}{}

\renewcommand{\J}{\mathbf{J}}
\renewcommand{\R}{\mathbb{R}}

\renewcommand{\d}{\mathbf{d}}

\renewcommand{\n}{\mathbf{n}}
\renewcommand{\p}{\mathbf{p}}
\renewcommand{\q}{\mathbf{q}}

\renewcommand{\t}{\mathbf{t}}
\renewcommand{\u}{\mathbf{u}}
\renewcommand{\v}{\mathbf{v}}
\renewcommand{\x}{\mathbf{x}}
\renewcommand{\y}{\mathbf{y}}
\renewcommand{\z}{\mathbf{z}}
\newcommand{\coarse}{\Omega'}
\newcommand{\fine}{\Omega}
\newcommand{\fineneural}{\Omega_{\theta}}
\newcommand{\map}{f}

\newcommand{\mapmlp}{g_{\theta}}
\newcommand{\mapinvmlp}{g_{\phi}}
\newcommand{\jac}{\J_{\map}}
\newcommand{\jacmlp}{\J_{\theta}}

\newcommand{\projection}{\mathcal{P}}
\newcommand{\normal}{\n_{\coarse}}
\newcommand{\tangent}{T_{\coarse}}
\newcommand{\tanone}{\t_{\coarse}^1}
\newcommand{\tantwo}{\t_{\coarse}^2}
\newcommand{\tangentfine}{T_{\theta}}
\newcommand{\tanonefine}{\t_{\theta}^1}
\newcommand{\tantwofine}{\t_{\theta}^2}
\newcommand{\normalfine}{\n_{\theta}}

\newcommand{\bijectiveloss}{\mathcal{L}_C(\theta, \phi)}
\newcommand{\projectionloss}{\mathcal{L}_P(\theta)}
\newcommand{\conformalloss}{\mathcal{L}_F(\theta)}

\usepackage{tabularx}
\usepackage{booktabs}
\usepackage{makecell}
\usepackage[table]{xcolor}
\definecolor{white}{rgb}{1,1,1}
\definecolor{lightbluishgrey}{rgb}{0.76471,0.84824,0.91647}

\usepackage{subcaption}
\usepackage{wrapfig}
\usepackage{graphicx}

\usepackage{listings}
\usepackage{layouts}
\makeatletter 
\newcommand{\layoutdetails}{%
\begin{tabular}{ll}
 \texttt{\textbackslash{textwidth}} & \printinunitsof{in}\prntlen{\textwidth} \\
\texttt{\textbackslash{linewidth}} & \printinunitsof{in}\prntlen{\linewidth} \\
Main text font &  \f@size pt \f@family \\
\sffamily \small Caption text font &  \sffamily \small \f@size pt \f@family \\
\end{tabular}%
}
\makeatother

%% file: sections/00_abstract.tex
Mesh processing pipelines are mature, but adapting them to newer non-mesh surface representations---which enable fast rendering with compact file size---requires costly meshing or transmitting bulky meshes, negating their core benefits for streaming applications.

We present a compact neural field that enables common geometry processing tasks across diverse surface representations. Given an input surface, our method learns a neural map from its coarse mesh approximation to the surface. The full representation totals only a few hundred kilobytes, making it ideal for lightweight transmission. Our method enables fast extraction of manifold and Delaunay meshes for intrinsic shape analysis, and compresses scalar fields for efficient delivery of costly precomputed results. Experiments and applications show that our fast, compact, and accurate approach opens up new possibilities for interactive geometry processing.

%% file: sections/01_introduction.tex
\section{Introduction}

\begin{figure}[b]
  \centering
  \vspace{-0.2cm}
  \includegraphics[width=1\columnwidth]{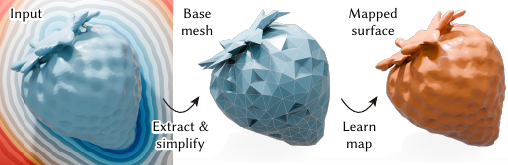}
  
  \caption{Given an input neural implicit surface, we extract a \emph{base mesh} that coarsely approximates the input surface, and learn a \emph{neural displacement field} that maps a base mesh to the input surface. }
  \label{fig:overview}
   \vspace{-0.4cm}
\end{figure}

Neural fields have emerged as powerful representations for visualizing 3D geometry. 
Neural Signed Distance Fields (SDFs) and Neural Radiance Fields (NeRFs) enable high-fidelity rendering at interactive frame rates \cite{Muller2022InstantNGP} while being significantly compact than traditional representations \cite{takikawa2023compactngp, li2022compressingvolumetricradiancefields}. The interactive speed and compactness are especially valuable for streaming applications like games and interactive apps, where the client must download data from the server and render it in real time.

Despite the impact of neural fields in rendering, the whole area of geometric computing spans beyond visualizing surfaces. In particular, intrinsic shape analysis, such as geodesic computation or Laplacian geometry processing, is central to applications like physics-based animation. However, these pipelines typically rely on manifold and Delaunay triangle meshes for robust processing. Applying them to neural surface representations requires mesh extraction: either on the client side, involving heavy remeshing algorithms, or on the server side, necessitating transmission of bulky meshes.

\begin{figure*}
  \centering
  \includegraphics[width=\linewidth]{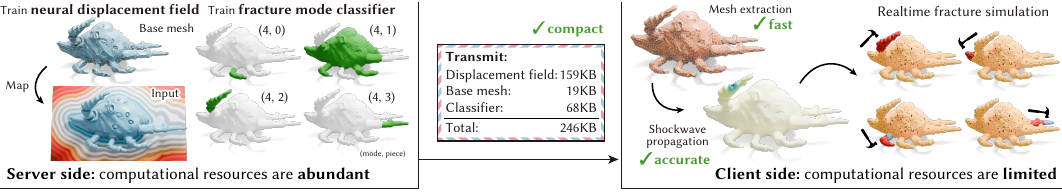}
  \caption{One application of our method is realtime fracture simulation \cite{Sellan2023BreakingGood}. 
  On the server side, we train the neural displacement field and the fracture mode classifier using our method. The neural networks are compact enough (246KB) to be sent to the client side, where we run mesh extraction and a sparse Cholesky factorization in under a second. This allows realtime fracture simulation that can be done in milliseconds.}
  \label{fig:fracture-modes}
\end{figure*}

We propose a method tailored for geometry processing in streaming applications, where server resources are abundant but client compute and bandwidth are limited.
Our goal is to minimize computational and transmission burdens on the client, while retaining high accuracy in geometry processing tasks.
To this end,  we learn a \emph{neural displacement field} on the server side that is both compact and fast to infer. This field maps a coarse mesh approximation---referred to as the \emph{base mesh}---to the input surface (Figure \ref{fig:overview}). Inspired by surface multigrid methods \cite{Liu2021SurfaceMultigrid, Gu2002GeometryImages, Lee1998maps}, the displacement field acts as a quasi one-to-one correspondence between the base mesh and the implicit surface. This enables fast, accurate extraction of manifold and Delaunay meshes on the client side.
Additionally, our method compresses scalar fields into compact neural representations, enabling heavy preprocessing to be offloaded to the server while keeping transmission overhead minimal.

Another key advantage of our method is its representation-agnostic design. It supports a wide range of shape representations: not only neural implicit surfaces, but also oriented point clouds and SDFs, with partial support for NeRFs, Gaussian splats, and non-oriented point clouds. This flexibility enables visualization using the original representation, while leveraging our extracted mesh for robust geometric computation.

Figure \ref{fig:fracture-modes} shows the overview of our pipeline where:
\begin{enumerate}
    \item the server side trains a neural displacement field and a scalar field that requires heavy computation,
    \item the compact MLPs and the base mesh are transmitted to the client side, and
    \item the client side subdivides the mesh and runs the simulation in an interactive runtime.
\end{enumerate}

We demonstrate that our fast, compact, and accurate method enables new types of interactive streaming applications on non-mesh surfaces, including geodesic computation, Laplacian-based processing, fracture simulation, surface sampling, and more.

%% file: sections/02_related-work.tex
\section{Related Work}

We discuss three alternative methods for our pipeline: mesh extraction on the client side, mesh extraction on the server side, and meshless methods.

\paragraph{Mesh extraction on the client side}

To perform geometry processing tasks on neural surfaces, one solution is to extract the triangle mesh on the client side. This approach can avoid transmitting bulky meshes over the internet. Although there are hundreds of optimization-based mesh extraction methods \cite{deAraujo2015SurveyImplicitSurface, huang2024surface, Berger2017SurveyofSurfaceRecon}, here we only focus on methods that do not require additional training or optimization on the client side, allowing mesh extraction in seconds.

Thanks to their robustness and simplicity, Marching Cubes \cite{Lorensen1987MarchingCubes} or Dual Contouring \cite{Ju2002DualContouring} or their variants \cite{Doi1991MarchingTets, Schaefer2004DualMarchingCubes, Lopes2003ImprovingMC, Schaefer2007ManifoldDC} often serve as the final component in surface reconstruction pipelines. Recently, several isosurfacing approaches have been proposed to leverage learned priors from a large dataset \cite{Chen2021NeuralMC, Chen2022NeuralDualContouring, sundararaman2024selfsuperviseddualcontouring}, while another method harnesses the behavior of the underlying implicit function \cite{Hwang2024ODC}. Also, Sharp and Crane \shortcite{Sharp2020LaplacianNonmanifold} find local connectivity of a point cloud with a simple proximity search.

\begin{figure}
  \centering
  \includegraphics[width=1\columnwidth]{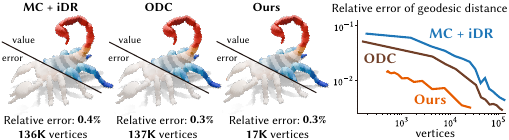}
  
  \caption{Our method creates a mesh that is accurate in computing geodesic distances with a significantly smaller number of vertices than isosurfacing methods \cite{Hwang2024ODC, Lorensen1987MarchingCubes}.  }

  \label{fig:mc-topology}
\end{figure}

However, most of these meshes have thin, poor-quality triangles,  inducing error in eigenvalue analysis (Figure \ref{fig:eigenvalues-mc}) or leading to longer runtime due to the excessive number of vertices (Figure \ref{fig:mc-topology}). Furthermore, many methods miss the local connectivity of the surface, leading to larger errors in geodesic computation (Figure \ref{fig:pointcloud-laplacian}). Finally, some isosurfacing methods (e.g., NDC \cite{Chen2022NeuralDualContouring}) produce non-manifold edges for most cases, making the numerical instability even worse.

\begin{figure}[b]
  \centering
  \includegraphics[width=\columnwidth]{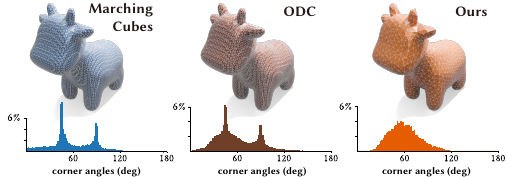}
  
  \caption{  
   Our mesh is free of thin sliver triangles unlike isosurfacing methods \cite{Lorensen1987MarchingCubes, Hwang2024ODC}, leading to accurate and efficient computation on the mesh.
  }
  \label{fig:mc-tessellation}
   
\end{figure}

Another tempting approach is to remesh bad quality meshes using, e.g., isotropic remeshers \cite{botsch2004remeshingapproach} or intrinsic Delaunay triangulation \cite{Sharp2019NavigatingIntrinsic}. However, the edge flip procedures in these methods are mostly $\mathcal{O}(n \log n)$, leading to notoriously long runtimes due to the excessive number of vertices in a bad quality mesh. Furthermore, such edge flip procedures are not easily parallelizable \cite{Cao2014GPUdelaunay}.

\begin{figure}
  \centering
  \includegraphics[width=\columnwidth]{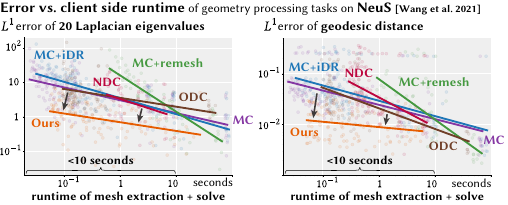}
  
  \caption{One alternative to our method is to extract the mesh on the client side using isosurfacing methods \protect\footnotemark, which has a trade-off between speed and accuracy. By additionally transmitting our representation, which is only 174KB in this setup, we extend the Pareto frontier towards the bottom left of the plot, especially where the runtime is below 10 seconds.}
  \label{fig:neus-client}
   
\end{figure}

In Figure \ref{fig:neus-client}, we show the time and error trade-off of the existing approaches. Raw isosurfacing methods (MC, ODC) or neural-powered methods (NDC) are fast but prone to higher error. Using an isotropic remesher \cite{botsch2004remeshingapproach} decreases the error sharply but suffers from a longer runtime to remesh. In contrast, our representation can extract manifold and Delaunay meshes with computationally cheap operations, leading to faster runtime with lower error.

\footnotetext{  \textbf{MC}: Marching Cubes \cite{Lorensen1987MarchingCubes}, \textbf{ODC}: Occupancy-Based Dual Contouring \cite{Hwang2024ODC}, \textbf{NDC}: Neural Dual Contouring \cite{Chen2022NeuralDualContouring}, 

\textbf{iDR}: intrinsic Delaunay refinement \cite{Sharp2019NavigatingIntrinsic}, \textbf{remesh}: isotropic remeshing \cite{botsch2004remeshingapproach}.}

\paragraph{Mesh extraction on the server side}

\begin{figure}
  \centering
  \includegraphics[width=1\columnwidth]{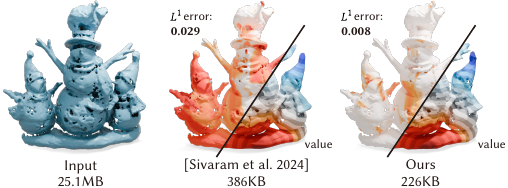}
  \caption{A state-of-the-art mesh compression method \cite{Sivaram2024NeuralGeometryFields} can compress the input mesh but cannot keep the accuracy of the geodesic distance. In contrast, our method achieves a 3.6x error reduction with a smaller file size.}
  
  \label{fig:ngf}
\end{figure}

Another solution is to extract the mesh on the server side and transmit it to the client. Since meshes are often bulky (Figure \ref{fig:filesize}), significant research has focused on mesh compression. Early approaches include lossless schemes \cite{Alliez2001ValenceDriven, deering1995geometrycompression, rossignac20013dcompression, szymczak2001edgebreaker, taubin1998geometriccompression, touma1998trianglemesh} and surface simplification techniques \cite{Garland1997SurfaceSimplification, surazhsky2003explicitsurface, szymczak2002piecewiseregular}.
Recently, neural methods have emerged, such as data-driven subdivision \cite{Chen2023NeuralProgressive, Liu2020NeuralSubdivision}, compression using differentiable rendering losses \cite{Sivaram2024NeuralGeometryFields}, and overfitting to a self-parametrization \cite{pentapati2025meshcompressionquantizedneural}. The latter two works are particularly relevant, as they also learn a neural displacement field from a coarse mesh to a fine one.
However, these methods are primarily designed to preserve visual appearance and often fail to retain local geometric connectivity, which is critical for intrinsic shape analysis (Figure \ref{fig:ngf}). Moreover, because they only accept mesh inputs, non-mesh representations must first be converted to meshes, introducing additional discretization error before compression even begins.

\begin{figure}
  \centering
  \includegraphics[width=1\columnwidth]{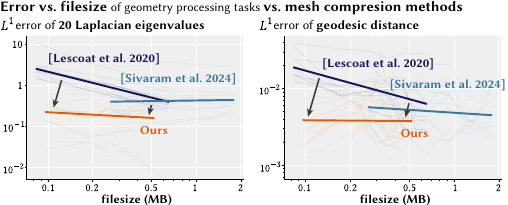}
  
  \caption{Another alternative to our method is to extract a mesh on the server side and transmit it to the client side, using surface simplification methods \cite{lescoat2020spectralmesh} or a state-of-the-art mesh compression method \cite{Sivaram2024NeuralGeometryFields}. We extend the Pareto frontier towards the bottom left of the plot, especially where the file size is below 500KB.}
  
  \label{fig:decimation}
   
\end{figure}

We argue that the true value of a mesh lies in its utility for computation. Triangle meshes provide a convenient finite-element basis for solving PDEs, and are supported by a rich ecosystem of software libraries (e.g., \textsc{CGAL} \cite{CGAL}, \textsc{libigl} \cite{libigl}). Prior work has explored mesh simplification tailored to preserve specific computational properties---for instance, Otaduy and Lin \shortcite{otaduy2003sensationpreserving} focused on preserving haptic features, while Lescoat et al. \shortcite{lescoat2020spectralmesh} targeted spectral characteristics. As shown in Figure \ref{fig:decimation}, our method significantly outperforms Lescoat et al. \shortcite{lescoat2020spectralmesh}, achieving lower geodesic and spectral error with smaller file sizes. This reinforces the idea that compact representations can still enable high-quality computation.

\paragraph{Meshless methods}

The third category circumvents the need for high-quality meshing and directly solves the problem on the surface.

Many works use neural networks to solve their geometry processing tasks, akin to our approach. Several early attempts try to derive the local properties like normal directions and curvatures on non-oriented point clouds \cite{Guerrero2018PCPNet, Ben-Shabat2019Nesti-Net, Pistilli2020PointCloudNormal,bednarik2020shape} or accurately derive these properties on neural SDFs \cite{chetan2023accuratedifferentialoperatorshybrid, Novello2022ExploringDifferentialGeometry}.
In order to support a broader set of tasks, several works \cite{Yang2021NeuralFields, williamson2024neuralgeometryprocessingspherical} use neural networks to represent a vector field on a surface and solve various geometry processing problems by regressing a loss. However, these works require additional training to run their task, making it infeasible to add interactive control on the client side (e.g., changing the position of the boundary condition). Other methods represent surface maps \cite{Morreale2021NeuralSurfaceMaps} or the Laplace-Beltrami operator \cite{Pang2024NeuralLaplacianOperator} using a neural network, but these works require the ground truth value computed on a discretized mesh, and little is said about inference time or compression.

Alternatively, Monte Carlo approaches \cite{Sawhney2020MonteCarloGeometry} or the closest point method (CPM) \cite{King2024ClosestPointMethod, li2023closest, Sugimoto2024ProjectedWoS} bypass the need for discretizing the boundary domain. However, naive Monte Carlo approaches require a precomputed conformal parametrization \citep[Section 6.5]{Sawhney2022Grid-freeMonteCarlo}, while CPM-based approaches fall short due to extremely slow computation on the client side (see Figure \ref{fig:cpm}).

%% file: sections/03_method.tex
\section{Method}

\begin{figure}
  \centering
  \includegraphics[width=1\columnwidth]{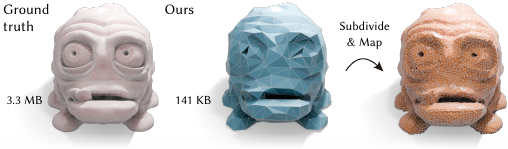}
  
  \caption{Our representation is compact and thus suitable for transmitting over the internet.}

  \label{fig:filesize}
   
\end{figure}

\begin{figure}
  \centering
  \includegraphics[width=1\columnwidth]{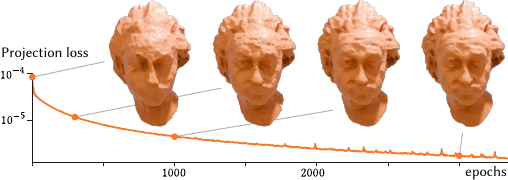}
  
  \caption{$\fineneural$ learns to approximate $\fine$ as training proceeds. }
  \label{fig:epoch-map}
  
\end{figure}

The input of our method is a target surface $\fine \subset \R^3$ represented as a non-mesh. Our output will be (1) a manifold base mesh $\coarse \subset \R^3$ that approximates the shape of $\fine$ and (2) a \emph{neural displacement field} $\mapmlp: \R^3 \mapsto \R^3$ parametrized by weights $\theta$, learned to approximate a map $\map: \coarse \mapsto \fine$.

To achieve this, we extract a base mesh using existing procedures (Section \ref{sec:coarse-approx}), coarsen it to a target face number $N_{\coarse}$ using QEM \cite{Garland1997SurfaceSimplification}, and finally learn $\mapmlp$. 
The compressibility relies on $N_{\coarse}$, and thus one may change this to trade off file size and accuracy. 
Also, we use a fairly small MLP to represent $\mapmlp$ (64 neurons with 2 hidden layers), allowing inference to be done in several milliseconds (Figure \ref{fig:injective-nets}).

We also use an intrinsic encoding scheme that further enhances the accuracy of our neural displacement field (Section \ref{sec:architecture}).
Using the base mesh $\coarse$ and the neural displacement field $\mapmlp$, one may extract the final mesh with cheap computation (Section \ref{sec:mesh-extraction}), or learn scalar fields on surfaces (Section \ref{sec:learning-scalar-fields}).

In what follows, let $\normal(\x) \in \R^3$ be the normal vector at $\x \in \coarse$, $\tangent(\x)$ be the tangent plane at $\x \in \coarse$, and $\tanone(\x), \tantwo(\x) \in \R^3$ be the orthogonal tangent vectors spanning $\tangent(\x)$. Please refer to the supplemental material for additional technical details.

\subsection{Querying Geometric Quantities}
\label{sec:geometric-queries}

The mapped surface $\mapmlp(\coarse) =: \fineneural$ has the same topology as $\coarse$, and as the training proceeds, $\fineneural$ approximates $\fine$ (see Figure \ref{fig:epoch-map}).
Before showing how to train the neural displacement field $\mapmlp$, let us explain how we can access to important geometric quantities on $\fineneural$, which our computation of $\mapmlp$ relies on.

\paragraph{Normals and tangent vectors}

\begin{wrapfigure}[7]{r}{0.44\columnwidth}

    \hspace{-14pt}
    \includegraphics[width=0.48\columnwidth]{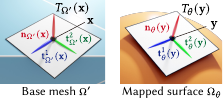}
    
    \label{subfig:jacobian}
  
\end{wrapfigure}
Given a point $\x \in \coarse$, one useful quantity on $\fineneural$ is the unit normal vector at $\y = \mapmlp(\x) \in \fineneural$ which we denote as $\normalfine(\y) \in \R^3$. With the normal vector $\normalfine(\y)$ in hand, we may also define the tangent plane $\tangentfine(\y)$ and two unit orthogonal vectors $\tanonefine(\y), \tantwofine(\y) \in \R^3$ spanning it (see inset).

Let
\begin{equation}
    \jacmlp(\x) = \frac{\partial}{\partial \x} \mapmlp(\x) \in \R^{3 \times 3}
\end{equation}
be the Jacobian of $\mapmlp$ at $\x$. Then, the tangent plane on $\y$ is spanned by two vectors $\jacmlp(\x) \tanone(\x)$ and $\jacmlp(\x) \tantwo(\x)$, leading to
\begin{equation}
    \normalfine(\y) = \frac{\jacmlp(\x) \tanone(\x) \times \jacmlp(\x) \tantwo(\x)}{\| \jacmlp(\x) \tanone(\x) \|_2 \| \jacmlp(\x) \tanone(\x) \|_2}.
\end{equation}

\paragraph{Local Jacobians}

Viewing $\mapmlp$ as a map from $\coarse$ to $\fineneural$, we can also define the \emph{pushforward} $d\mapmlp^{\x}: \tangent(\x) \mapsto \tangentfine(\mapmlp(\x))$ which defines how the local neighborhood at $\tangent(\x)$ is mapped to the local neighborhood at $\tangentfine(\mapmlp(\x))$. The pushforward can be described as a mapping from a local coordinate space spanned by $\tanone, \tantwo$ to one spanned by $\tanonefine, \tantwofine$, where the transformation can be given as a $2 \times 2$ matrix
\begin{equation}
    \label{eq:jacobian}
    \jac(\x) = \begin{bmatrix}
        \tanonefine(\mapmlp(\x)), \tantwofine(\mapmlp(\x))
    \end{bmatrix}^{T} 
    \jacmlp(\x)
    \begin{bmatrix}
        \tanone(\x), \tantwo(\x)
    \end{bmatrix}
\end{equation}
which we call the \emph{local Jacobian}  matrix.

\subsection{Loss Function}
\label{sec:learning-map}

Our loss function contains two terms, which can all be computed using the same $s$ samples $\x_1, ..., \x_s \in \coarse$ sampled on the mesh $\coarse$. In order to integrate the loss on $\fineneural$ instead of $\coarse$, we draw them such that the distribution of the points on $\coarse$ is proportional to $\left| \det \jac(\x) \right|$ using rejection sampling (Figure \ref{fig:sampling}). See the supplemental material for other optional losses and training details.

To make our neural displacement field work as a quasi 1-to-1 mapping between the base mesh and the fine surface, we also learn the inverse neural map $\mapinvmlp$ parametrized by weights $\phi$. The inverse map is only used for training, and thus does not need to be sent to the client side.

\begin{figure}
  \centering
  \includegraphics[width=1\columnwidth]{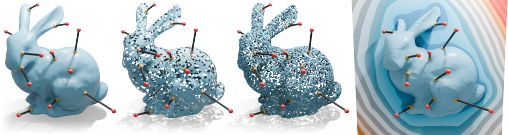}
  
  \caption{Our projection operator $\projection$ supports various surface representations, such as (left to right) meshes, polygon soups, point clouds, or neural occupancy functions where the Eikonal equation is not necessarily met (right).  }
    
  \label{fig:projection}
   
\end{figure}

\paragraph{Projection loss}

The first loss ensures that the mapped surface $\fineneural$ matches the input surface $\fine$. In order to compute the loss, the surface representation of $\fine$ must be able to define a \emph{projection operator} $\projection(\p)$ (Figure \ref{fig:projection}) that takes a point $\p \in \R^3$ as input and returns
\begin{equation}
    \label{eq:projection}
    \projection(\p) = 
    \begin{cases}
        ~\p & (\p \in \fine),\\
        ~\q \in \fine& (\p \notin \fine).
    \end{cases}
\end{equation}
We will discuss later in Section \ref{sec:various-rep} on how to perform this projection on various surface representations.

We want $\projection(\mapmlp(\x))$ to match $\mapmlp(\x)$ for every $\x \in \coarse$, and thus we add the following loss
\begin{equation}
    \projectionloss = \frac{1}{s} \sum_{i=1}^{s} \| \mapmlp(\x_i) - \projection \left(\mapmlp(\x_i) \right) \|_2^2.
\end{equation}

\paragraph{Cycle consistency loss}

The second loss encourages $\mapinvmlp$ to be the inverse of $\mapmlp$. This can be encouraged by adding the following loss
\begin{equation}
    \bijectiveloss = \frac{1}{s} \sum_{i=1}^{s} \| \mapinvmlp(\mapmlp(\x_i)) - \x_i \|_2^2.
\end{equation}

\begin{wrapfigure}[7]{r}{0.44\columnwidth}
    \vspace{-12pt}
  
    \hspace{-14pt}
    \includegraphics[width=0.48\columnwidth]{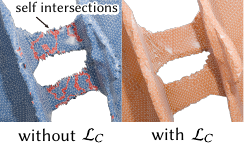}
    
    \label{subfig:cyclic-consistency}
  
\end{wrapfigure}
We also found that this loss contributes to having a smoother surface free of self-intersections (see inset).

We opt for regressing a loss to encourage cycle consistency in a weak sense. This design choice is because existing injective neural networks are either not expressive enough \cite{behrmann2019invertibleresidualnetworks} or slow 
 \cite{sun2024tuttenet} (see Figure \ref{fig:injective-nets}).

\paragraph{Total loss}

Given all the losses at hand, we now define our final optimization problem
\begin{equation}
    \label{eq:total-loss}
    \underset{\theta, \phi}{\mathrm{argmin}}~ \lambda_C \bijectiveloss  + \lambda_P \projectionloss,
\end{equation}
where $\lambda_C, \lambda_P$ are hyperparameters.

\subsection{Neural Network Architecture}
\label{sec:architecture}

We model the residual of our neural displacement field $\mapmlp(\x) - \x$ and its inverse $\mapinvmlp(\x) - \x$ using two MLPs with 2 hidden layers, 64 internal hidden units, and ReLU activation functions. We also applied periodic positional encoding \cite{mildenhall2020nerf} with 8 layers.

For the neural displacement field $\mapmlp$, we also apply an intrinsic encoding that involves learnable $d$-dimensional feature vectors assigned to each vertex. This helps to learn scalar or vector fields on the surface with finer details (see Figure \ref{fig:transmit-result}). The idea itself is similar to previous mesh compression works \cite{Sivaram2024NeuralGeometryFields}, but ours is a more general representation that extends naturally to base triangle meshes instead of quad meshes.

Suppose that a point $\x_i \in \coarse$ lies on a face that contains vertices $\v_i^1, \v_i^2, \v_i^3 \in \mathbb{R}^3$ on the base mesh. For each vertex, we have a learnable $d$-dimensional feature vector  $\u_i^1, \u_i^2, \u_i^3 \in \mathbb{R}^3$. Let $[w_i^1, w_i^2, w_i^3]$ be the barycentric coordinate of $\x_i$,

we interpolate the feature vectors as
\begin{equation}
    \u_i = w_i^1 \u_i^1 + w_i^2 \u_i^2 + w_i^3 \u_i^3 \in \mathbb{R}^d
\end{equation}
and feed this vector to the MLP in addition to the positional encoding vectors.

\subsection{Mesh Extraction}
\label{sec:mesh-extraction}

\begin{figure}
  \centering
  \includegraphics[width=1\columnwidth]{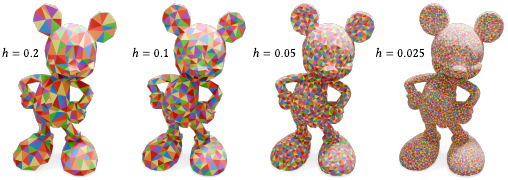}
  
  \caption{Our neural displacement field can progressively refine the mesh, allowing users to trade off runtime and accuracy in their computation tasks.  }
  \label{fig:progressive}
   
\end{figure}

Given the neural displacement field $\mapmlp$, one may extract a mesh that is suitable for geometry processing. We achieve this by subdividing the base mesh and mapping all the vertices to the target surface $\fineneural$ using $\mapmlp$.

Given a user-specified parameter $h$ (Figure \ref{fig:progressive}), we subdivide the edge on the base mesh $\coarse$ at the midpoint if the mapped edge is longer than $h$, and remove one of its incident vertices if it is shorter than $h/4$. In order to retain the geometry of the base mesh, we allow vertex removal only for the ones that were added during midpoint subdivision. We also conduct (extrinsic) edge flips if it reduces the ODT energy \cite{chen2004optimal}, which is the sum of the squared edge lengths weighted by its incident triangle areas. To keep the bijection between the base mesh and the mapped mesh, we only allow edge flips if the edge can be flipped intrinsically (see Sharp et al. \shortcite{Sharp2019NavigatingIntrinsic} Section 4.1).
We iterate these three procedures 5 times, and after that we run intrinsic Delaunay triangulation \cite{Sharp2019NavigatingIntrinsic} to ensure the Delaunay condition. As our base mesh extraction always creates manifold meshes, the subdivided mesh is also guaranteed to be manifold.

This mesh can be used for geometry processing tasks such as computing geodesic distances or Laplacian geometry processing. The solution computed on the mesh can be queried by simply taking the closest point query from the fine surface to the subdivided mesh and barycentrically interpolating the per-vertex value. This can be combined with rendering algorithms that get the explicit ray intersection position with the surface (e.g., sphere tracing).

Akin to progressive physics simulation methods \cite{zhang2024progressivedynamics}, one may also progressively refine the mesh to have a \emph{preview} at earlier runtime and a high-quality result later (Figure \ref{fig:progressive}). This allows the client application to trade off the computational budget and accuracy based on their demand.

\subsection{Learning Scalar Fields}
\label{sec:learning-scalar-fields}

Given the trained neural displacement field $\mapmlp$, we may also train a neural network to represent a scalar field on the surface $\fine$. This is especially useful when one wants to compress a precomputed scalar field that is expensive to compute on the client side, like the object's fracture modes (Figure \ref{fig:fracture-modes}).

Let $u: \fine \mapsto \mathbb{R}$ be a scalar field computed on the mesh of $\fine$. We overfit a neural network $u_{\varphi}: \coarse \mapsto \mathbb{R}$ such that $u_{\varphi}$ approximates $u \circ \mapmlp$. As $u_{\varphi}$ is a scalar field defined on the coarse mesh $\coarse$, we use the encoding scheme we introduced in Section \ref{sec:architecture} to parametrize the function intrinsically to the coarse mesh.

Let $\x_1, ..., \x_s$ be uniformly sampled points on the mesh approximating $\Omega$.
If $u$ is a continuous function, we regress a loss
\begin{equation}
    \mathcal{L_{\varphi}} = \frac{1}{s} \sum_{i}^s \| u_{\varphi}(\x_i) - u(\mapmlp(\x_i)) \|^2.
\end{equation}
Likewise, if $u$ gives a binary value, we regress
\begin{equation}
    \mathcal{L_{\varphi}} = \frac{1}{s} \sum_{i}^s \mathcal{L}_{\mathrm{BCE}}(u_{\varphi}(\x_i), u(\mapmlp(\x_i)),
\end{equation}
where $\mathcal{L}_{\mathrm{BCE}}(x, y) = 
y\,\ln\bigl(1/(1+e^{-x})\bigr)
\;+\;(1-y)\,\ln\bigl(e^{-x}/(1+e^{-x}) $ is the binary cross entropy loss.\

%% file: sections/04_shape-representations.tex
\section{Supporting Various Representations}
\label{sec:various-rep}

\begin{table}
  \caption{We support various surface representations.}
  \label{tab:representations}
  \small

  \centering
  \footnotesize
  \begin{tabular}{p{0.13\textwidth}p{0.15\textwidth}p{0.15\textwidth}}
    \toprule
    
    \makecell[l]{Surface \\ Representation} & 
    \makecell[l]{(1) Base mesh $\coarse$ \\ $\blacktriangleright$ Section \ref{sec:coarse-approx}} & 
    \makecell[l]{(2) Projection $\projection(\p, \d)$ \\ $\blacktriangleright$ Section \ref{sec:projection}} \\
    \midrule

    Analytic SDFs & Marching Cubes & SDF projection \\

    \rowcolor{ACMOrange!30}
    Grid-interpolated implicits & Marching Cubes & Newton's method \\

    Neural implicits/NeuS & Marching Cubes & Newton's method \\

    \rowcolor{ACMOrange!30}
    Polygon soups/ \hfill\hfill \linebreak Oriented point clouds & Marching Cubes on the Generalized Winding Number (GWN) field \cite{Barill2018FWN} & Ray tracing the GWN field using Harnack Tracing \cite{Gillespie2024RayTracingHarmonic} \\

    NeRFs (partial support) & Marching Cubes on the TSDF \cite{zhang2022nerfusionfusingradiancefields}  & Ray tracing the density field \\

    \rowcolor{ACMOrange!30}
    Non-oriented point \hfill\hfill \linebreak clouds/Gaussian splats \linebreak (partial support)& Dilation 
    & Closest point query \\
    \bottomrule
  \end{tabular}
  
\end{table}

Our method requires the surface representation to be able to (1) extract the base mesh $\coarse$ and (2) define a projection operator $\projection$. We outline the supported surface representations in Table~\ref{tab:representations}.  

\subsection{Extracting Base Mesh $\coarse$}
\label{sec:coarse-approx}

Most surface representations support inside/outside occupancy functions, like \textbf{analytic SDFs}, \textbf{grid-
interpolated
implicits}, \textbf{neural
implicits} \cite{park2019deepsdflearningcontinuoussigned}, \textbf{NeuS} \cite{wang2021neus}, and the Generalized Winding Number \cite{Jacobson2013RobustInside, Barill2018FWN} for \textbf{oriented point clouds}. For such cases, we apply Marching Cubes \cite{Lorensen1987MarchingCubes} on an upsampled grid and coarsen it with surface simplification algorithms \cite{Garland1997SurfaceSimplification}.

\textbf{NeRFs}, \textbf{Non-oriented point clouds}, or \textbf{3D Gaussian splats} do not have a definitive boundary of the occupancy function, and we thus only claim partial support. However, one may still use an existing method that estimates the occupancy (e.g., TSDF \cite{zhang2022nerfusionfusingradiancefields} for NeRFs) or dilate a point cloud and coarsen it (Figure \ref{fig:dilation}). We also note that several neural representations have a well-behaved occupancy function for novel view synthesis data (e.g., NeuS \cite{wang2021neus}).

\subsection{Projection Operator $\projection$}
\label{sec:projection}

Our second requirement for $\fine$ is the ability to define the projection operator (Eq.~\ref{eq:projection}).

\paragraph{Implicit functions}
For shapes represented as the zero level set of an implicit function $\phi: \mathbb{R}^3 \mapsto \mathbb{R}$ including \textbf{neural implicits}, \textbf{NeuS}, and \textbf{grid-interpolated implicits}, we use Newton's method that undergoes an iterative procedure 
\begin{equation}
    \p_{i+1} = \p_i - \phi(\p_i) \frac{\nabla \phi(\p_i)}{\| \nabla \phi(\p_i) \|^2}, ~\p_0 = \p,
\end{equation}
which is also used for sampling level sets on neural implicits \cite{Atzmon2019ControllingNeural}.

\paragraph{Closest point queries}
A closest point query to $\fine$ can be directly used as our $\projection$, an operator commonly required in representation-agnostic PDE solvers \cite{Sawhney2020MonteCarloGeometry, King2024ClosestPointMethod}. For \textbf{analytic SDFs} $\phi_S: \mathbb{R}^3 \mapsto \mathbb{R}$, the projection of $\p$ is obtained by $\p - \phi_S(\p) \nabla \phi_S(\p)$ \cite{Marschner2023ConstructiveSolid}, which we call SDF projection.

For \textbf{non-oriented point clouds} and \textbf{3D Gaussian splats} sampled on $\fine$, we take the closest point in the point cloud or the center positions of the Gaussians (see Figure \ref{fig:gs}).

For \textbf{oriented point clouds} or \textbf{polygon soups}, we use Harnack tracing \cite{Gillespie2024RayTracingHarmonic} and terminate the ray where the Generalized Winding Number (GWN) equals $0.5$ \cite{Jacobson2013RobustInside}. To get the ray direction, we apply a Gaussian kernel to the normal vectors of the eight nearest neighbors of $\p$ in the point cloud and negate them if the GWN is smaller than $0.5$. This is similar in spirit to the vector field definition in Poisson Surface Reconstruction \citep[Section 4.2]{kazhdan2006poisson}. These adjustments will provide a more accurate estimate of the closest point to the surface than a simple nearest neighbor search (see the supplemental material for details). We note that this projection operator may not converge, and in such a case, we simply omit the nonconvergent samples and only use samples that succeed.

\paragraph{Ray traceable geometry}

For \textbf{Neural Radiance Fields (NeRFs)} \cite{mildenhall2020nerf}, we shoot two rays $\p + t \n_{\theta}(\p)$ and $\p - t \n_{\theta}(\p)$ to get the closest intersection point, where $\n_{\theta}(\p)$ is the normal direction of $\fineneural$ at $\p$. The ray intersection is where the density function equals a threshold (Figure \ref{fig:nerf}). In our implementation, we naively sample points uniformly along the ray to find the first intersection.

%% file: sections/20_experiments.tex
\section{Results}

In this section, we report the core results of our method in terms of speed, compressibility, accuracy, and versatility.
Please see the supplemental material for other details such as baseline setup, ablations, hyperparameters, and extensions.

\subsection{Experiments and Ablations}
\label{sec:exp}

\paragraph{Mesh extraction}

Our mesh extraction tackles the multiobjective problem regarding (1) runtime, (2) file size, and (3) error of computation tasks (geodesics/eigenvalue analysis). As there are three axes, it is difficult to project all baseline methods on a single 2D plot and do a fair comparison. We thus show three versions of plots: comparison with fast mesh extraction methods (Figure \ref{fig:neus-client}), with mesh compression methods (Figure \ref{fig:decimation}), and a 3D plot with all of the axes (see the attached HTML file in the supplemental material).

Ideally, one would want shorter runtime, compact file size, and lower error in their computation tasks. In our plots, we show that our method pushes the Pareto frontier towards this goal. For example, our method outperforms existing mesh extraction methods in regions where the runtime is below 10 seconds (Figure \ref{fig:neus-client}) . Likewise,  our method outperforms existing mesh compression methods in the range of -500KB (Figure \ref{fig:decimation}).

We also qualitatively show that our method can extract a compact mesh with lower error (Figures \ref{fig:mc-topology}, \ref{fig:mc-tessellation}), preserves the spectral pattern of the ground truth surface (Figures \ref{fig:eigenvalues-mc}), and has a compact representation compared to mesh compression methods (Figure \ref{fig:ngf}).

\paragraph{Other representations}

In Figure \ref{fig:pointcloud}, we show that our method extends well for oriented point clouds, showing the same runtime-reducing effect as in neural implicits. We also qualitatively show that our method can preserve the spectral pattern of the surface for point clouds (Figure \ref{fig:pointcloud-laplacian}).

\paragraph{Learning scalar fields}

To validate the performance of our scalar field learning algorithm in Section \ref{sec:learning-scalar-fields}, we learned the smallest 10 eigenfunctions of a shape to a single neural network by registering them to the latent code (Figure \ref{fig:transmit-result}). We compared our method to a baseline MLP that uses the same parameters (2 hidden layers, 32 neurons, 8 positional encoding dimensions) but takes a 3D point on the fine surface as input and outputs a function value. By using our intrinsic encoding with $d=8$, we achieved 3.6x lower error compared to the baseline, without increasing the number of neurons of the MLP.

\subsection{Applications}
\label{sec:app}

The accurate connectivity of our generated mesh can accommodate applications that require manifold and Delaunay meshes, such as vector field design \cite{Knoppel2013GloballyOptimal} or stripe patterns on surfaces \cite{Knoppel2015StripePatterns} (Figure \ref{fig:artistic}). Having a good-quality mesh also contributes to interactively editing the design properties, like the heat source positions for heat simulation (Figure \ref{fig:teaser}).

Our method can be used for realtime fracture simulation \cite{Sellan2023BreakingGood} (Figure \ref{fig:fracture-modes}). In their method, fracture mode decomposition takes minutes, which we want to avoid running on the client side. Using our intrinsic encoding, we train a binary classifier that tells whether a point $\p \in \fine$ belongs to a certain piece of a certain fracture mode. Thanks to this compact classifier (68KB) and our efficient mesh extraction, the client side only required 0.6 seconds of precomputation, allowing us to run the fracture simulation in 7 ms.

Our method can handle cases where the shape of the base mesh and the final geometry are significantly different, like the torus and Bob (Figure \ref{fig:same-topology}). Given several pairs of anchor points to guide the training, our neural displacement field can successfully learn the mapping (see the supplemental for details). This can allow swapping the texture of the two surfaces simultaneously on the client side \cite{Schmidt2023SurfaceMaps}, but without sending the full mesh.

Finally, one may run different strategies to sample points on the mapped surface (Figure \ref{fig:sampling}), which could be useful for visualization. As the determinant of our local Jacobian $\det \J_f$ is proportional to the distortion of our map, we achieve white noise sampling by uniformly generating samples on the base mesh and accepting samples at a rate proportional to it. Likewise, we achieve blue noise sampling by slightly modifying Poisson disk sampling \cite{Bridson2007PoissonDisk} to evaluate the radius of the annulus on the mapped points instead of on the base mesh.

%% file: sections/80_conclusion.tex
\section{Conclusion and Discussion}

We propose a representation-agnostic method that learns a neural displacement field from a base mesh to the input surface.
Our method addresses the requirements for streaming applications to be fast, compact, and accurate.

The cost of our fast and compact mesh extraction is its low visual quality. This is due to our design choice to prioritize speed and accuracy in intrinsic shape analysis, where Delaunay and non-manifold triangles matter more than their extrinsic appearance. We thus suppose our representation to be used as an \emph{add-on} to existing representations that already excel in visual quality.

One remaining fundamental problem is how to visualize deformation or fracture on neural surfaces. To just visualize it, one may sample point clouds on the fine surface using our method (Figure \ref{fig:sampling}), but this cannot leverage the existing neural rendering pipeline. One potential solution is to distort the rays accordingly to the deformation \cite{Seyb2019NonlinearSphere, sun2024tuttenet}, but it is unclear how to generalize them to various surface representations.

Our method does not guarantee cycle consistency strictly, but applying recent flow matching methods \cite{lipman2024flowmatchingguidecode} could potentially ensure it.

Our method only supports mesh extraction for surface meshes, but many physics-based animation tasks work better with volumetric tetrahedral meshes. Thus, fast extraction of them using neural networks will be an exciting future direction.

Finally, we hope our method can empower the computer graphics community and beyond to run their interactive mesh processing applications on non-meshes. We plan to make our code public upon acceptance.

%% file: sections/89_figureonly.tex
\clearpage

\begin{figure}
  \centering
  \includegraphics[width=1\columnwidth]{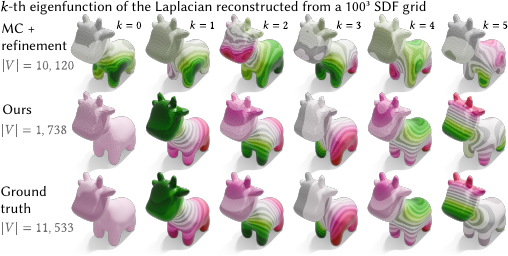}
  
  \caption{Meshes obtained by Marching Cubes \cite{Lorensen1987MarchingCubes} suffer from thin sliver triangles, leading to catastrophic failure in analyzing intrinsic properties even after intrinsic Delaunay refinement \cite{Sharp2019NavigatingIntrinsic}. In contrast, meshes obtained using our method qualitatively preserve the pattern even with a significantly smaller number of vertices.  }

  \label{fig:eigenvalues-mc}
\end{figure}

\begin{figure}
  \centering
  \includegraphics[width=1\columnwidth]{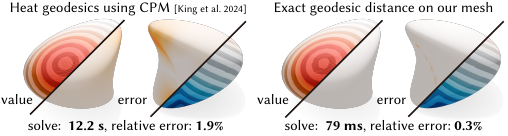}
  
  \caption{Running the Closest Point Method \cite{King2024ClosestPointMethod} on the client side suffers from not only longer computation time but also larger error. Ours, while requiring training on the server side, is significantly faster and more accurate on the client side.}
  \label{fig:cpm}
   
\end{figure}

\begin{figure}
  \centering
  \includegraphics[width=1\columnwidth]{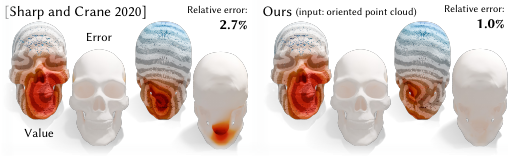}
  
  \caption{Unlike the point cloud Laplacian of Sharp and Crane \shortcite{Sharp2020LaplacianNonmanifold}, our mesh can distinguish the connectivity of thin parts, leading to accurate geodesic computation.}
  \label{fig:nonmanifold-laplacian}
   
\end{figure}

\begin{figure}[htbp]
  \centering
  \includegraphics[width=1\columnwidth]{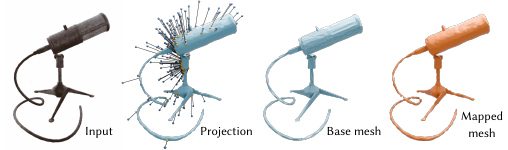}
  \caption{Our method partially supports Neural Radiance Fields \cite{mildenhall2020nerf}. The projection is done by shooting a ray towards a direction $\d$ and getting its first hit point with the levelset of the density function.}
  \label{fig:nerf}

\end{figure}

\begin{figure}
  \centering
  \includegraphics[width=1\columnwidth]{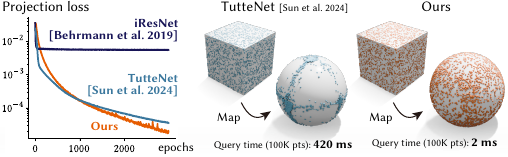}
  
  \caption{Invertible neural networks  \cite{behrmann2019invertibleresidualnetworks} do not have enough expressivity to learn even the simplest map from a cube to a sphere (left). TutteNet \cite{sun2024tuttenet} can learn this map, but the map induces large distortions which makes mesh extraction infeasible, and the query time is slow (middle).}
  \label{fig:injective-nets}
  
\end{figure}

\begin{figure}
  \centering
  \includegraphics[width=1\columnwidth]{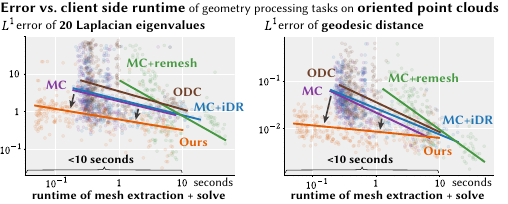}
  \caption{Our method also shows the time- and error-reducing effect for oriented point clouds. Comparison methods were run on the fast generalized winding number \cite{Barill2018FWN}.}
  \label{fig:pointcloud}
\end{figure}

\begin{figure}[htbp]
  \centering
  \includegraphics[width=1\columnwidth]{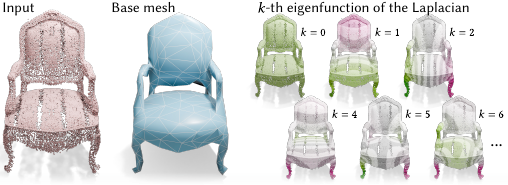}
  
  \caption{Our method partially supports 3D Gaussian Splats \cite{Kerbl20233DGaussianSplat} by treating them as a volumetric point cloud. The mapped mesh is clean enough to have a positive definite Laplacian, accommodating applications like eigenmode analysis. }
  \label{fig:gs}
   
\end{figure}

\begin{figure}
  \centering
  \includegraphics[width=1\columnwidth]{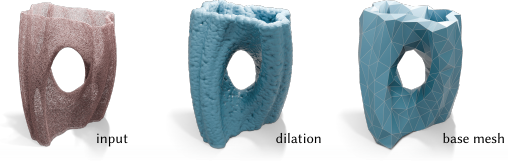}
  
  \caption{On a non-oriented point cloud where the occupancy function is unavailable, one may dilate the points for a certain radius by adding an offset to the distance function on a regular grid. }
  \label{fig:dilation}
   
\end{figure}

\clearpage

\begin{figure}[htbp]
  \centering
  \includegraphics[width=1\columnwidth]{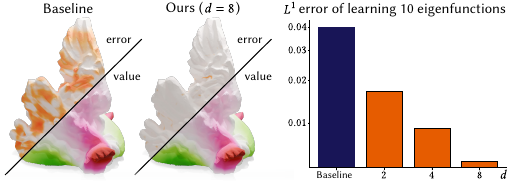}
  
  \caption{Using our map and our intrinsic encoding, one may train a neural network that represents a scalar function on a surface, like the eigenfunctions of the Laplacian (left). By increasing the dimension of the feature vectors $d$, our method can achieve lower error without changing the number of layers/neurons of the MLP (right).}
  \label{fig:transmit-result}
   
\end{figure}

\begin{figure}
  \centering
  \includegraphics[width=1\columnwidth]{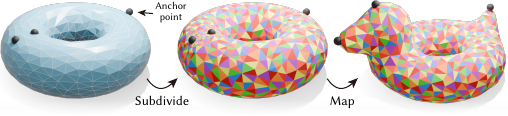}
  \caption{Our method allows base meshes with different shapes but with the same topology. }
  
  \label{fig:same-topology}
\end{figure}

\begin{figure}
  \centering
  \includegraphics[width=1\linewidth]{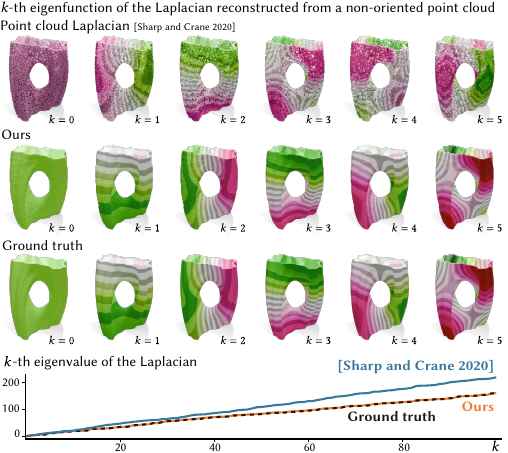}
  \caption{On a thin shell, the spectral pattern of the point cloud Laplacian \cite{Sharp2020LaplacianNonmanifold} significantly differs from the ground truth. In contrast, our mesh successfully reconstructs its spectral pattern. }
  
  \label{fig:pointcloud-laplacian}

\end{figure}

\begin{figure}
  \centering
  \includegraphics[width=1\columnwidth]{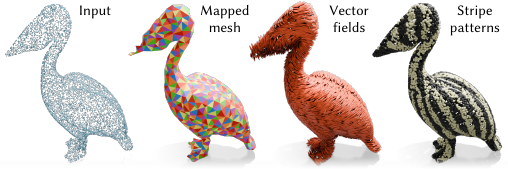}
  \caption{One may use our map for visualizing vector fields \cite{Knoppel2013GloballyOptimal} or stripe patterns \cite{Knoppel2015StripePatterns} on the client side.}
  
  \label{fig:artistic}
\end{figure}

\begin{figure}
  \centering
  \includegraphics[width=1\columnwidth]{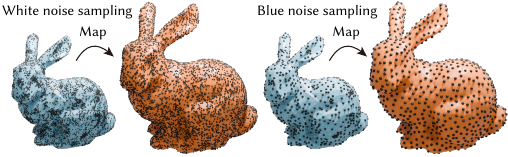}
  \caption{Our map can facilitate different surface sampling strategies like white noise and blue noise sampling.}
  \label{fig:sampling}
\end{figure}

%% file: sections/90_appendix.tex
\section{Implementation}
\label{sec:impl}

\subsection{Training}
\label{sec:train}

As a preprocessing step, we uniformly scale the non-mesh surface $\fine$ and the base mesh $\coarse$ such that $\coarse$ is bounded by a cube $[-1.5, 1.5]^3$.
Before training for the main loss, we initialize the neural networks using Kaiming initialization \cite{Kaiming2015}. Next, we train $\mapmlp$ and $\mapinvmlp$ so that they represent the identity. For both models, we sample $q = 32,768$ points $\z_1, ..., \z_q \in \coarse^3$ and train the loss
\begin{equation}
    \label{eq:initialization-loss}
    \underset{\theta}{\mathrm{argmin}}~ \sum_{i=1}^q \| \mapmlp(\z_i) - \z_i \|_2^2
\end{equation}
by running Adam \cite{kingma2017adam} for 1,000 epochs ($\theta$ and $\phi$ are interchangeable here).
After initialization, we train our loss by running Adam for 10,000 epochs. We set the sample number $s = 32,768$ and resample every ten epochs.

\subsection{Code Base}

We implemented our training code in \textsc{Python} using \textsc{pytorch} \cite{pytorch} for standard operations on MLPs and auto differentiation, and \textsc{libigl} \cite{libigl} and \textsc{gpytoolbox} \cite{gpytoolbox} for standard geometry processing routines. 

We implemented our remeshing algorithm using \textsc{geometry central} \cite{geometrycentral} for storing the intrinsic triangulation in the integer coordinate data structure \cite{Gillespie2021IntegerCoordinates}. 

\section{Baseline Setup}

We discuss the baseline setups of the experiments we show in the main text.
All baseline experiments were run on a desktop computer with an NVIDIA GeForce RTX 3070 and 8GB of RAM.

\subsection{Implementation}

We implemented geodesic computation and Laplacian eigenmode analysis on \textsc{c++}  using \textsc{geometry central} \cite{geometrycentral} and called it from \textsc{python} using \textsc{pybind11} \cite{pybind11}. For geodesic computation we used the MMP algorithm \cite{mitchell1987mmp}.

In all of our experiments, we computed the geodesic distance from the closest point from $[1, 1, 1]^T$ and $[-1,-1,-1]^T$ to the mesh, and the 20 smallest eigenvalues of the Laplace-Beltrami operator. We evaluate the $L^1$ error of geodesic distance by querying the closest point from each vertex on the ground truth mesh to the extracted mesh, and then applying barycentric interpolation to get the value on that closest point.
We used the same code in the core algorithm that computes the geodesic distance or the spectral decomposition.

The Laplace-Beltrami operator is positive semidefinite only if the triangulation is manifold and Delaunay \cite{wardetzky2007nofreelunch}. Thus, to avoid numerical instability for meshes that do not satisfy them, we shift the eigenvalues with a small value $\sigma = 10^{-3}$.

Also, many methods do not guarantee topological identity to the ground truth; for example, Marching Cubes can easily create outliers that are not connected to other parts of the surface, leading the geodesic distances of those parts to infinity. Thus, we extract the triangles that are singly connected before we run the tasks.

\subsection{Isosurfacing Methods (Figure 6)}

We compared our method to Neural Dual Contouring (NDC) \cite{Chen2021NeuralMC}, Occupancy-Based Dual Contouring (ODC) \cite{Hwang2024ODC}, and Marching Cubes (MC) \cite{Lorensen1987MarchingCubes} to measure the runtime vs. error. All the runtimes include (1) the query time of the neural network, (2) the time required for mesh extraction, (3) the time to run geodesic computation or eigenmode analysis, and (4) the time to query the solution on the ground truth vertex positions. 
We observed that the runtime of (4) was mostly in milliseconds and was fairly negligible.

For NDC and ODC, we used the official code release by the authors, while for MC, we used a GPU implementation on \textsc{warp} \cite{warp2022}. For NDC we used NDCx, a version that gives higher reconstruction accuracy. We ran each of these isosurfacing methods with a grid size of $h=0.01, 0.02, ..., 0.1$, where the object fits in a bounding box of $[-1.5, 1.5]^3$. We note that NDC timed out for all the inputs with $h=0.01$, and we simply omitted those samples from the experiment.

For our method, we extracted the mesh with a target edge length $h=0.01, 0.014, 0.025, 0.035, 0.05, 0.07, 0.1, 0.14, 0.2$. 

We also ran additional meshing after MC, using either intrinsic Delaunay refinement (iDR) \cite{Sharp2019NavigatingIntrinsic} or isotropic remeshing \cite{botsch2004remeshingapproach}. For iDR we used the code provided by the authors in \textsc{geometry central}, while for isotropic remeshing, we used the \textsc{c++} binding implementation from \textsc{gpytoolbox}.

In this experiment, we used NeuS \cite{wang2021neus} using the implementation of \textsc{sdfstudio} \cite{Yu2022SDFStudio} trained on the DTU dataset \cite{aanaes2016large}.
To get the ground truth, we extracted the mesh with Marching Cubes \cite{Lorensen1987MarchingCubes} on a $300^3$ grid and then applied isotropic remeshing \cite{botsch2004remeshingapproach} with $h=0.005$.

\paragraph{Fast winding numbers}

We also used the same setup to compute the isosurface of the generalized winding number (GWN) \cite{Barill2018FWN}, which we used their implementation on \textsc{libigl} to query the field.

All the abovementioned methods except NDC were runnable. NDC only accepts signed distance functions, which makes our measurement infeasible. Although NDC does provide a point cloud version (UNDC), we found that UNDC causes out of memory on a 10K point cloud, while their \texttt{--noisypc} option that divides point clouds into batches takes minutes to finish.

\subsection{Mesh Compression Methods (Figure 8)}

We compared the error vs. file size to Spectral Mesh Simplification (SMS) \cite{lescoat2020spectralmesh}, a decimation algorithm that can preserve the spectral property well, and Neural Geometry Fields for Meshes (NGF) \cite{Sivaram2024NeuralGeometryFields}, a state-of-the-art mesh compression method. For SMS, we measured the theoretical optimal file size of a mesh given by $(32 \times 3 \times |V| + 3 \times |F| \times \log_2{|V|}) / 8$ bytes, where $|V|$ is the number of vertices and $|F|$ is the number of faces. For NGF \cite{Sivaram2024NeuralGeometryFields}, we measured the size of the binary file they produce, while for our neural displacement field, we measured the file size of our weights, feature vectors, and base mesh saved in a torch script file.

For both of the baseline methods, we used the source code provided by the authors. For SMS, we set the number of eigenvalues to preserve to 20 and changed the number of target vertices, and for NGF, we changed the number of vertices of the base mesh.

For our method, we changed the target face number $N_{\coarse} = 2000, 2800, 4000, ..., 32000$ with $d$ proportional to the reciprocal of $N_{\coarse}$, and plotted the smallest error found by changing $h$.

In this experiment, we randomly selected 10 models from \textsc{thingi10k} \cite{zhou2016thingi10kdataset100003dprinting} that have a file size of over 1MB and the whole mesh is singly connected. For all three methods (SMS, NGF, ours), we treated the mesh as a pure SDF and did not draw any information other than the signed distance to the mesh. For SMS and NGF, we extracted the mesh with Marching Cubes \cite{Lorensen1987MarchingCubes} on a $300^3$ grid and then applied isotropic remeshing \cite{botsch2004remeshingapproach} with $h=0.005$, and gave this mesh as an input.

\subsection{3D Plot (supplemental HTML file)}

In order to perform a comparison with all of the mesh extraction and mesh compresison methods, we plot all the three axes (data file size, runtime on the client side, error) on a single 3D plot. In the plot, we show that our method breaks the tradeoff between the two classes of methods, and leads to a interesting new region of the plot. Please open the HTML file for details. 

In the plot, we used the same setup as Figure 6 where we compared to mesh extraction methods on NeuS trained on the DTU dataset. For additional data points for mesh compression methods, we fed the mesh we treated as the ground truth ($300^3$ grid Marching Cubes + isotropic remeshing) to SMS. Although mesh extraction methods do not require any additional data transfer, we plotted them on the plane of file size 100KB for clarity, which is a file size that any of the methods including ours did not achieve.

We note that NGF \cite{Sivaram2024NeuralGeometryFields} relies on the \emph{Tri to Quad by smart triangle pairing} filter in \textsc{meshlab} \cite{meshlab},  but we found out that this procedure is not robust enough on high genus surfaces. We observed that NGF fails to compress meshes on 43 out of 105 test meshes. Thus, in order to keep a fair comparison, we omitted NGF from this plot.

\subsection{Learning Scalar Fields (Figure 20)}

In this experiment, we learned the 10 eigenfunctions $u_1, ..., u_10: \fine \mapsto \mathbb{R}$ into a single MLP $u_{\xi}$. This MLP takes the 3D point and the number of the eigenfunction $k < 10$, and outputs the (learned) $k$-th eigenfunction at that point. The baseline is an MLP that takes four dimensions (xyz and the latent code) as input and has 2 layers, 32 neurons, and 8 positional encoding layers. We compare this baseline with a neural network that uses our intrinsic encoding with feature vector dimensions $d=2, 4, 8$.

We trained the MLPs using a loss
\begin{equation}
    \frac{1}{s} \sum_k^{10} \sum_{i=1}^s \| u_{\xi}(\p_i, k) - u_k(\p_i) \|_2^2,
\end{equation}
with $s=3276$ samples, and iterated Adam \cite{kingma2017adam} for 40,000 epochs with a learning rate $10^{-3}$.

\section{Ablations and Measurements}

\subsection{Feature Vector Dimension $d$}

We show in Figure \ref{fig:error_layer} that the final loss decreases as we increase the feature vector dimension $d$.

\subsection{Cycle Consistency}
\label{sec:cycle-consistency}

Our method encourages cycle consistency only by a soft penalty, and hence it cannot be strictly guaranteed. However, existing neural network structures that guarantee cycle consistency are not expressive and fast enough for our purpose. We report that the average $ \bijectiveloss $ of the final epoch when trained on 34 models from \textsc{thingi10k} \cite{zhou2016thingi10kdataset100003dprinting} was $4.8 \times 10^{-7}$, which was sufficiently small enough for our purpose.

\subsection{Projection Operators}

\begin{figure}
  \centering
  \includegraphics[width=1\columnwidth]{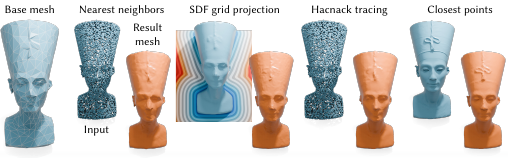}
  
  \caption{
 Having a better apporoximate of the closest point to the surface leads to higher accuracy.}
  \label{fig:various-representations}
   
\end{figure}

\begin{figure}
    \centering
    \includegraphics[width=0.5\columnwidth]{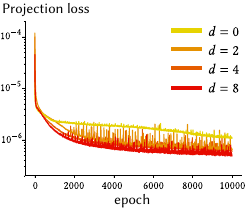}
    
    \caption{The projection loss decreases as we increase the feature vector dimension $d$.}
    \label{fig:error_layer}
  
\end{figure}

In Section 4.2 in the main text, we discussed different projection operators that approximates the closest point query.
We observed that the visual quality of the extracted mesh highly depends on the accuracy of the closest point query (see Figure \ref{fig:various-representations}). In other words, the closer the projection operator becomes to give the closest point, the more accurate the final outcome.

\begin{table}
  \caption{We report the training runtime for each surface representation. }
  \vspace{-0.3cm}
  \label{tab:runtime}
  \begin{tabular}{lrrrrr}
    \toprule
    Surface representation & Runtime  \\
    \midrule
    Mesh treated as an SDF & 18 minutes \\
    NeuS & 32 minutes \\
    Oriented point cloud & 7 hours 12 minutes \\

    \bottomrule
  \end{tabular}
  \vspace{-0.1cm}
\end{table}

\subsection{Runtime}

We report the runtime of our training in Table \ref{tab:runtime}. As we used a relatively low-budget desktop computer with NVIDIA GeForce RTX 3070 and only 8GB of RAM, this could potentially be sped up on a stronger one.

\subsection{Hyperparameters}

The hyperparameters of our training are $\lambda_C, \lambda_P$ and the learning rate. We report the choice of our hyperparameters in Table \ref{tab:time}.

\begin{table}
  \caption{We report the hyperparameters $\lambda_C, \lambda_P$, and the learning rate on different representations.}
  \vspace{-0.3cm}
  \label{tab:time}
  \begin{tabular}{lrrrrr}
    \toprule
    Surface representation & $\lambda_C $ & $ \lambda_P$ &  learning rate \\
    \midrule
    Mesh treated as SDFs & $10^2$ & $10^4$ & $1\times10^{-3}$ \\
    
    Neural implicits / NeuS & $10^2$ & $10^4$ & $1\times10^{-3}$ \\

    Non-oriented point clouds & $10^5$ & $10^4 $ & $1\times10^{-3}$\\
    Oriented point clouds & $10^4$ & $10^4 $ & $1\times10^{-3}$ \\
    NeRFs & $10^4$ & $10^4 $& $1\times10^{-3}$ \\
    3D Gaussian splats & $10^5$ & $10^4 $  & $1\times10^{-3}$ \\

    \bottomrule
  \end{tabular}
  \vspace{-0.1cm}
\end{table}

\section{Additional Losses}

We provide additional losses that can be added to our vanilla loss during our training.

\subsection{Anchor Point Loss}

To guide the training of the map in Figure 17 in the main text, we set a pair of anchor points $\p = \{\p_1, ..., \p_m\} \in \coarse$ 
and $\q = \{\q_1, ..., \q_m\} \in \fine$ such that $\mapmlp(\p_i)$ matches $\q_i$. 
This can be done by adding a loss
\begin{equation}
    \lambda_A \sum_{i=1}^{m} \| \mapmlp(\p_i) - \mapmlp(\q_i) \|^2
\end{equation}
which we set $\lambda_A = 0.1 \lambda_P$.

\subsection{Normal Alignment Loss}

In order to improve the visual quality of the mapped surface, one may encourage the normal direction $\normalfine(\mapmlp(\x_i))$ on $\mapmlp(\x_i) \in \fineneural$ to align with the normal direction $\n_{\fine}(\y_\fine(\x_i))$ on the projected point $\y_\fine(\x_i) = \projection(\mapmlp(\x), \normalfine(\mapmlp(\x_i))) \in \fine$ by adding a loss
\begin{equation}
    \mathcal{L}_N(\theta) = \frac{1}{s} \sum_{i=1}^s \left\| \normalfine(\mapmlp(\x_i)) - \n_{\fine}(\y_\fine(\x_i)) \right\|^2.
\end{equation}

\subsection{Local Conformality Loss}

In order to obtain a map $\map$ with cycle consistency, it must be free of \emph{foldovers}, i.e., points where the local Jacobian $\jac(\x)$ is negative or indefinite \cite{Garanzha2021FoldoverFree}. Also, the map $\map$ should not have large anisotropic distortions if the user wants to use our map for, e.g., texture mapping. Thus, the user may wish it to be as conformal as possible.

A typical approach to encourage foldover-free and conformal maps \cite{hormann2000mips, garanzha2000barrier} is to use a barrier-like loss
\begin{equation}
    \label{eq:barrier-injective}
    \sum_{i=1}^s \frac{\mathrm{tr}~ \J_i^{T}\J_i}{\det~ \J_i} ,
\end{equation}
where $\J_i = \jac(\x_i)$.
However, our map is prone to having negative Jacobians during the training, leading this loss to approach infinity. Another tempting approach is to wrap $\det \J_i$ with a function that always gives a positive value \cite{Garanzha2021FoldoverFree}, but this still caused instabilities due to orders of magnitudes of larger losses for negative Jacobians.

To solve these issues, we instead use a softer penalty loss
\begin{equation}
    \conformalloss = \frac{1}{s} \sum_{i=1}^s \chi \left(- \frac{\det~ \J_i}{\mathrm{tr}~ \J_i^{T}\J_i }\right),
    \label{eq:conformality}
\end{equation}
\begin{wrapfigure}[5]{r}{0.4\columnwidth}
  \vspace{-5pt}
    \hspace{-15pt}
  
    \includegraphics[width=0.45\columnwidth]{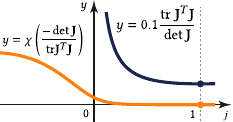}
    \label{subfig:monotonicity}
  
\end{wrapfigure}
where $\chi(t) = \frac{1}{10} \log(1 + \exp(10 t))$ is the $\mathrm{SoftPlus}$ function. The $\J_i$ that gives the global minimum of Eq. \ref{eq:conformality} also gives the global minimum for Eq. \ref{eq:barrier-injective}; in the inset we show the case where $\J = \begin{bmatrix}
    1 & 0 \\
    0 & j
\end{bmatrix}$.

%% file: zzz_main_arxiv.bbl

\begin{thebibliography}{94}


\ifx \showCODEN    \undefined \def \showCODEN     #1{\unskip}     \fi
\ifx \showDOI      \undefined \def \showDOI       #1{#1}\fi
\ifx \showISBNx    \undefined \def \showISBNx     #1{\unskip}     \fi
\ifx \showISBNxiii \undefined \def \showISBNxiii  #1{\unskip}     \fi
\ifx \showISSN     \undefined \def \showISSN      #1{\unskip}     \fi
\ifx \showLCCN     \undefined \def \showLCCN      #1{\unskip}     \fi
\ifx \shownote     \undefined \def \shownote      #1{#1}          \fi
\ifx \showarticletitle \undefined \def \showarticletitle #1{#1}   \fi
\ifx \showURL      \undefined \def \showURL       {\relax}        \fi
\providecommand\bibfield[2]{#2}
\providecommand\bibinfo[2]{#2}
\providecommand\natexlab[1]{#1}
\providecommand\showeprint[2][]{arXiv:#2}

\bibitem[Aan{\ae}s et~al\mbox{.}(2016)]%
        {aanaes2016large}
\bibfield{author}{\bibinfo{person}{Henrik Aan{\ae}s}, \bibinfo{person}{Rasmus~Ramsb{\o}l Jensen}, \bibinfo{person}{George Vogiatzis}, \bibinfo{person}{Engin Tola}, {and} \bibinfo{person}{Anders~Bjorholm Dahl}.} \bibinfo{year}{2016}\natexlab{}.
\newblock \showarticletitle{Large-Scale Data for Multiple-View Stereopsis}.
\newblock \bibinfo{journal}{\emph{International Journal of Computer Vision}} (\bibinfo{year}{2016}), \bibinfo{pages}{1--16}.
\newblock


\bibitem[Alliez and Desbrun(2001)]%
        {Alliez2001ValenceDriven}
\bibfield{author}{\bibinfo{person}{Pierre Alliez} {and} \bibinfo{person}{Mathieu Desbrun}.} \bibinfo{year}{2001}\natexlab{}.
\newblock \showarticletitle{Valence-Driven Connectivity Encoding for 3D Meshes}.
\newblock \bibinfo{journal}{\emph{Computer Graphics Forum}} \bibinfo{volume}{20}, \bibinfo{number}{3} (\bibinfo{year}{2001}), \bibinfo{pages}{480--489}.
\newblock
\urldef\tempurl%
\url{https://doi.org/10.1111/1467-8659.00541}
\showDOI{\tempurl}
\showeprint{https://onlinelibrary.wiley.com/doi/pdf/10.1111/1467-8659.00541}


\bibitem[Atzmon et~al\mbox{.}(2019)]%
        {Atzmon2019ControllingNeural}
\bibfield{author}{\bibinfo{person}{Matan Atzmon}, \bibinfo{person}{Niv Haim}, \bibinfo{person}{Lior Yariv}, \bibinfo{person}{Ofer Israelov}, \bibinfo{person}{Haggai Maron}, {and} \bibinfo{person}{Yaron Lipman}.} \bibinfo{year}{2019}\natexlab{}.
\newblock \bibinfo{booktitle}{\emph{Controlling neural level sets}}.
\newblock \bibinfo{publisher}{Curran Associates Inc.}, \bibinfo{address}{Red Hook, NY, USA}.
\newblock


\bibitem[Barill et~al\mbox{.}(2018)]%
        {Barill2018FWN}
\bibfield{author}{\bibinfo{person}{Gavin Barill}, \bibinfo{person}{Neil~G. Dickson}, \bibinfo{person}{Ryan Schmidt}, \bibinfo{person}{David I.~W. Levin}, {and} \bibinfo{person}{Alec Jacobson}.} \bibinfo{year}{2018}\natexlab{}.
\newblock \showarticletitle{Fast winding numbers for soups and clouds}.
\newblock \bibinfo{journal}{\emph{ACM Trans. Graph.}} \bibinfo{volume}{37}, \bibinfo{number}{4}, Article \bibinfo{articleno}{43} (\bibinfo{date}{jul} \bibinfo{year}{2018}), \bibinfo{numpages}{12}~pages.
\newblock
\showISSN{0730-0301}
\urldef\tempurl%
\url{https://doi.org/10.1145/3197517.3201337}
\showDOI{\tempurl}


\bibitem[Bednarik et~al\mbox{.}(2020)]%
        {bednarik2020shape}
\bibfield{author}{\bibinfo{person}{Jan Bednarik}, \bibinfo{person}{Shaifali Parashar}, \bibinfo{person}{Erhan Gundogdu}, \bibinfo{person}{Mathieu Salzmann}, {and} \bibinfo{person}{Pascal Fua}.} \bibinfo{year}{2020}\natexlab{}.
\newblock \showarticletitle{Shape reconstruction by learning differentiable surface representations}. In \bibinfo{booktitle}{\emph{Proceedings of the IEEE/CVF Conference on Computer Vision and Pattern Recognition}}. \bibinfo{pages}{4716--4725}.
\newblock


\bibitem[Behrmann et~al\mbox{.}(2019)]%
        {behrmann2019invertibleresidualnetworks}
\bibfield{author}{\bibinfo{person}{Jens Behrmann}, \bibinfo{person}{Will Grathwohl}, \bibinfo{person}{Ricky T.~Q. Chen}, \bibinfo{person}{David Duvenaud}, {and} \bibinfo{person}{Jörn-Henrik Jacobsen}.} \bibinfo{year}{2019}\natexlab{}.
\newblock \bibinfo{title}{Invertible Residual Networks}.
\newblock
\newblock
\showeprint[arxiv]{1811.00995}~[cs.LG]
\urldef\tempurl%
\url{https://arxiv.org/abs/1811.00995}
\showURL{%
\tempurl}


\bibitem[Ben-Shabat et~al\mbox{.}(2019)]%
        {Ben-Shabat2019Nesti-Net}
\bibfield{author}{\bibinfo{person}{Yizhak Ben-Shabat}, \bibinfo{person}{Michael Lindenbaum}, {and} \bibinfo{person}{Anath Fischer}.} \bibinfo{year}{2019}\natexlab{}.
\newblock \showarticletitle{Nesti-Net: Normal Estimation for Unstructured 3D Point Clouds Using Convolutional Neural Networks}. In \bibinfo{booktitle}{\emph{2019 IEEE/CVF Conference on Computer Vision and Pattern Recognition (CVPR)}}. \bibinfo{pages}{10104--10112}.
\newblock
\urldef\tempurl%
\url{https://doi.org/10.1109/CVPR.2019.01035}
\showDOI{\tempurl}


\bibitem[Berger et~al\mbox{.}(2017)]%
        {Berger2017SurveyofSurfaceRecon}
\bibfield{author}{\bibinfo{person}{Matthew Berger}, \bibinfo{person}{Andrea Tagliasacchi}, \bibinfo{person}{Lee~M. Seversky}, \bibinfo{person}{Pierre Alliez}, \bibinfo{person}{Gaël Guennebaud}, \bibinfo{person}{Joshua~A. Levine}, \bibinfo{person}{Andrei Sharf}, {and} \bibinfo{person}{Claudio~T. Silva}.} \bibinfo{year}{2017}\natexlab{}.
\newblock \showarticletitle{A Survey of Surface Reconstruction from Point Clouds}.
\newblock \bibinfo{journal}{\emph{Computer Graphics Forum}} \bibinfo{volume}{36}, \bibinfo{number}{1} (\bibinfo{year}{2017}), \bibinfo{pages}{301--329}.
\newblock
\urldef\tempurl%
\url{https://doi.org/10.1111/cgf.12802}
\showDOI{\tempurl}
\showeprint{https://onlinelibrary.wiley.com/doi/pdf/10.1111/cgf.12802}


\bibitem[Botsch and Kobbelt(2004)]%
        {botsch2004remeshingapproach}
\bibfield{author}{\bibinfo{person}{Mario Botsch} {and} \bibinfo{person}{Leif Kobbelt}.} \bibinfo{year}{2004}\natexlab{}.
\newblock \showarticletitle{A remeshing approach to multiresolution modeling}. In \bibinfo{booktitle}{\emph{Proceedings of the 2004 Eurographics/ACM SIGGRAPH Symposium on Geometry Processing}} (Nice, France) \emph{(\bibinfo{series}{SGP '04})}. \bibinfo{publisher}{Association for Computing Machinery}, \bibinfo{address}{New York, NY, USA}, \bibinfo{pages}{185–192}.
\newblock
\showISBNx{3905673134}
\urldef\tempurl%
\url{https://doi.org/10.1145/1057432.1057457}
\showDOI{\tempurl}


\bibitem[Bridson(2007)]%
        {Bridson2007PoissonDisk}
\bibfield{author}{\bibinfo{person}{Robert Bridson}.} \bibinfo{year}{2007}\natexlab{}.
\newblock \showarticletitle{Fast Poisson disk sampling in arbitrary dimensions}. In \bibinfo{booktitle}{\emph{ACM SIGGRAPH 2007 Sketches}} (San Diego, California) \emph{(\bibinfo{series}{SIGGRAPH '07})}. \bibinfo{publisher}{Association for Computing Machinery}, \bibinfo{address}{New York, NY, USA}, \bibinfo{pages}{22–es}.
\newblock
\showISBNx{9781450347266}
\urldef\tempurl%
\url{https://doi.org/10.1145/1278780.1278807}
\showDOI{\tempurl}


\bibitem[Cao et~al\mbox{.}(2014)]%
        {Cao2014GPUdelaunay}
\bibfield{author}{\bibinfo{person}{Thanh-Tung Cao}, \bibinfo{person}{Ashwin Nanjappa}, \bibinfo{person}{Mingcen Gao}, {and} \bibinfo{person}{Tiow-Seng Tan}.} \bibinfo{year}{2014}\natexlab{}.
\newblock \showarticletitle{A GPU accelerated algorithm for 3D Delaunay triangulation}. In \bibinfo{booktitle}{\emph{Proceedings of the 18th Meeting of the ACM SIGGRAPH Symposium on Interactive 3D Graphics and Games}} (San Francisco, California) \emph{(\bibinfo{series}{I3D '14})}. \bibinfo{publisher}{Association for Computing Machinery}, \bibinfo{address}{New York, NY, USA}, \bibinfo{pages}{47–54}.
\newblock
\showISBNx{9781450327176}
\urldef\tempurl%
\url{https://doi.org/10.1145/2556700.2556710}
\showDOI{\tempurl}


\bibitem[CGAL(2024)]%
        {CGAL}
CGAL \bibinfo{year}{2024}\natexlab{}.
\newblock \bibinfo{title}{{CGAL, Computational Geometry Algorithms Library}}.
\newblock
\newblock
\newblock
\shownote{\url{http://www.cgal.org}}.


\bibitem[Chen and Xu(2004)]%
        {chen2004optimal}
\bibfield{author}{\bibinfo{person}{Long Chen} {and} \bibinfo{person}{Jin-chao Xu}.} \bibinfo{year}{2004}\natexlab{}.
\newblock \showarticletitle{Optimal delaunay triangulations}.
\newblock \bibinfo{journal}{\emph{Journal of Computational Mathematics}} (\bibinfo{year}{2004}), \bibinfo{pages}{299--308}.
\newblock


\bibitem[Chen et~al\mbox{.}(2023)]%
        {Chen2023NeuralProgressive}
\bibfield{author}{\bibinfo{person}{Yun-Chun Chen}, \bibinfo{person}{Vladimir Kim}, \bibinfo{person}{Noam Aigerman}, {and} \bibinfo{person}{Alec Jacobson}.} \bibinfo{year}{2023}\natexlab{}.
\newblock \showarticletitle{Neural Progressive Meshes}. In \bibinfo{booktitle}{\emph{ACM SIGGRAPH 2023 Conference Proceedings}} (Los Angeles, CA, USA) \emph{(\bibinfo{series}{SIGGRAPH '23})}. \bibinfo{publisher}{Association for Computing Machinery}, \bibinfo{address}{New York, NY, USA}, Article \bibinfo{articleno}{84}, \bibinfo{numpages}{9}~pages.
\newblock
\showISBNx{9798400701597}
\urldef\tempurl%
\url{https://doi.org/10.1145/3588432.3591531}
\showDOI{\tempurl}


\bibitem[Chen et~al\mbox{.}(2022)]%
        {Chen2022NeuralDualContouring}
\bibfield{author}{\bibinfo{person}{Zhiqin Chen}, \bibinfo{person}{Andrea Tagliasacchi}, \bibinfo{person}{Thomas Funkhouser}, {and} \bibinfo{person}{Hao Zhang}.} \bibinfo{year}{2022}\natexlab{}.
\newblock \showarticletitle{Neural dual contouring}.
\newblock \bibinfo{journal}{\emph{ACM Trans. Graph.}} \bibinfo{volume}{41}, \bibinfo{number}{4}, Article \bibinfo{articleno}{104} (\bibinfo{date}{July} \bibinfo{year}{2022}), \bibinfo{numpages}{13}~pages.
\newblock
\showISSN{0730-0301}
\urldef\tempurl%
\url{https://doi.org/10.1145/3528223.3530108}
\showDOI{\tempurl}


\bibitem[Chen and Zhang(2021)]%
        {Chen2021NeuralMC}
\bibfield{author}{\bibinfo{person}{Zhiqin Chen} {and} \bibinfo{person}{Hao Zhang}.} \bibinfo{year}{2021}\natexlab{}.
\newblock \showarticletitle{Neural marching cubes}.
\newblock \bibinfo{journal}{\emph{ACM Trans. Graph.}} \bibinfo{volume}{40}, \bibinfo{number}{6}, Article \bibinfo{articleno}{251} (\bibinfo{date}{Dec.} \bibinfo{year}{2021}), \bibinfo{numpages}{15}~pages.
\newblock
\showISSN{0730-0301}
\urldef\tempurl%
\url{https://doi.org/10.1145/3478513.3480518}
\showDOI{\tempurl}


\bibitem[Chetan et~al\mbox{.}(2023)]%
        {chetan2023accuratedifferentialoperatorshybrid}
\bibfield{author}{\bibinfo{person}{Aditya Chetan}, \bibinfo{person}{Guandao Yang}, \bibinfo{person}{Zichen Wang}, \bibinfo{person}{Steve Marschner}, {and} \bibinfo{person}{Bharath Hariharan}.} \bibinfo{year}{2023}\natexlab{}.
\newblock \bibinfo{title}{Accurate Differential Operators for Hybrid Neural Fields}.
\newblock
\newblock
\showeprint[arxiv]{2312.05984}~[cs.CV]
\urldef\tempurl%
\url{https://arxiv.org/abs/2312.05984}
\showURL{%
\tempurl}


\bibitem[Cignoni et~al\mbox{.}(2008)]%
        {meshlab}
\bibfield{author}{\bibinfo{person}{Paolo Cignoni}, \bibinfo{person}{Marco Callieri}, \bibinfo{person}{Massimiliano Corsini}, \bibinfo{person}{Matteo Dellepiane}, \bibinfo{person}{Fabio Ganovelli}, {and} \bibinfo{person}{Guido Ranzuglia}.} \bibinfo{year}{2008}\natexlab{}.
\newblock \showarticletitle{{MeshLab: an Open-Source Mesh Processing Tool}}. In \bibinfo{booktitle}{\emph{Eurographics Italian Chapter Conference}}, \bibfield{editor}{\bibinfo{person}{Vittorio Scarano}, \bibinfo{person}{Rosario~De Chiara}, {and} \bibinfo{person}{Ugo Erra}} (Eds.). \bibinfo{publisher}{The Eurographics Association}.
\newblock
\showISBNx{978-3-905673-68-5}
\urldef\tempurl%
\url{https://doi.org/10.2312/LocalChapterEvents/ItalChap/ItalianChapConf2008/129-136}
\showDOI{\tempurl}


\bibitem[de~Ara\'{u}jo et~al\mbox{.}(2015)]%
        {deAraujo2015SurveyImplicitSurface}
\bibfield{author}{\bibinfo{person}{B.~R. de Ara\'{u}jo}, \bibinfo{person}{Daniel~S. Lopes}, \bibinfo{person}{Pauline Jepp}, \bibinfo{person}{Joaquim~A. Jorge}, {and} \bibinfo{person}{Brian Wyvill}.} \bibinfo{year}{2015}\natexlab{}.
\newblock \showarticletitle{A Survey on Implicit Surface Polygonization}.
\newblock \bibinfo{journal}{\emph{ACM Comput. Surv.}} \bibinfo{volume}{47}, \bibinfo{number}{4}, Article \bibinfo{articleno}{60} (\bibinfo{date}{May} \bibinfo{year}{2015}), \bibinfo{numpages}{39}~pages.
\newblock
\showISSN{0360-0300}
\urldef\tempurl%
\url{https://doi.org/10.1145/2732197}
\showDOI{\tempurl}


\bibitem[Deering(1995)]%
        {deering1995geometrycompression}
\bibfield{author}{\bibinfo{person}{Michael Deering}.} \bibinfo{year}{1995}\natexlab{}.
\newblock \showarticletitle{Geometry compression}. In \bibinfo{booktitle}{\emph{Proceedings of the 22nd Annual Conference on Computer Graphics and Interactive Techniques}} \emph{(\bibinfo{series}{SIGGRAPH '95})}. \bibinfo{publisher}{Association for Computing Machinery}, \bibinfo{address}{New York, NY, USA}, \bibinfo{pages}{13–20}.
\newblock
\showISBNx{0897917014}
\urldef\tempurl%
\url{https://doi.org/10.1145/218380.218391}
\showDOI{\tempurl}


\bibitem[Doi and Koide(1991)]%
        {Doi1991MarchingTets}
\bibfield{author}{\bibinfo{person}{Akio Doi} {and} \bibinfo{person}{Akio Koide}.} \bibinfo{year}{1991}\natexlab{}.
\newblock \showarticletitle{An Efficient Method of Triangulating Equi-Valued Surfaces by Using Tetrahedral Cells}.
\newblock \bibinfo{journal}{\emph{IEICE TRANSACTIONS on Information}} \bibinfo{volume}{E74-D}, \bibinfo{number}{1} (\bibinfo{date}{January} \bibinfo{year}{1991}), \bibinfo{pages}{214--224}.
\newblock


\bibitem[Garanzha(2000)]%
        {garanzha2000barrier}
\bibfield{author}{\bibinfo{person}{VA Garanzha}.} \bibinfo{year}{2000}\natexlab{}.
\newblock \showarticletitle{The barrier method for constructing quasi-isometric grids}.
\newblock \bibinfo{journal}{\emph{Computational Mathematics and Mathematical Physics}}  \bibinfo{volume}{40} (\bibinfo{year}{2000}), \bibinfo{pages}{1617--1637}.
\newblock


\bibitem[Garanzha et~al\mbox{.}(2021)]%
        {Garanzha2021FoldoverFree}
\bibfield{author}{\bibinfo{person}{Vladimir Garanzha}, \bibinfo{person}{Igor Kaporin}, \bibinfo{person}{Liudmila Kudryavtseva}, \bibinfo{person}{Fran\c{c}ois Protais}, \bibinfo{person}{Nicolas Ray}, {and} \bibinfo{person}{Dmitry Sokolov}.} \bibinfo{year}{2021}\natexlab{}.
\newblock \showarticletitle{Foldover-free maps in 50 lines of code}.
\newblock \bibinfo{journal}{\emph{ACM Trans. Graph.}} \bibinfo{volume}{40}, \bibinfo{number}{4}, Article \bibinfo{articleno}{102} (\bibinfo{date}{jul} \bibinfo{year}{2021}), \bibinfo{numpages}{16}~pages.
\newblock
\showISSN{0730-0301}
\urldef\tempurl%
\url{https://doi.org/10.1145/3450626.3459847}
\showDOI{\tempurl}


\bibitem[Garland and Heckbert(1997)]%
        {Garland1997SurfaceSimplification}
\bibfield{author}{\bibinfo{person}{Michael Garland} {and} \bibinfo{person}{Paul~S. Heckbert}.} \bibinfo{year}{1997}\natexlab{}.
\newblock \showarticletitle{Surface simplification using quadric error metrics}. In \bibinfo{booktitle}{\emph{Proceedings of the 24th Annual Conference on Computer Graphics and Interactive Techniques}} \emph{(\bibinfo{series}{SIGGRAPH '97})}. \bibinfo{publisher}{ACM Press/Addison-Wesley Publishing Co.}, \bibinfo{address}{USA}, \bibinfo{pages}{209–216}.
\newblock
\showISBNx{0897918967}
\urldef\tempurl%
\url{https://doi.org/10.1145/258734.258849}
\showDOI{\tempurl}


\bibitem[Gillespie et~al\mbox{.}(2021)]%
        {Gillespie2021IntegerCoordinates}
\bibfield{author}{\bibinfo{person}{Mark Gillespie}, \bibinfo{person}{Nicholas Sharp}, {and} \bibinfo{person}{Keenan Crane}.} \bibinfo{year}{2021}\natexlab{}.
\newblock \showarticletitle{Integer coordinates for intrinsic geometry processing}.
\newblock \bibinfo{journal}{\emph{ACM Trans. Graph.}} \bibinfo{volume}{40}, \bibinfo{number}{6}, Article \bibinfo{articleno}{252} (\bibinfo{date}{Dec.} \bibinfo{year}{2021}), \bibinfo{numpages}{13}~pages.
\newblock
\showISSN{0730-0301}
\urldef\tempurl%
\url{https://doi.org/10.1145/3478513.3480522}
\showDOI{\tempurl}


\bibitem[Gillespie et~al\mbox{.}(2024)]%
        {Gillespie2024RayTracingHarmonic}
\bibfield{author}{\bibinfo{person}{Mark Gillespie}, \bibinfo{person}{Denise Yang}, \bibinfo{person}{Mario Botsch}, {and} \bibinfo{person}{Keenan Crane}.} \bibinfo{year}{2024}\natexlab{}.
\newblock \showarticletitle{Ray Tracing Harmonic Functions}.
\newblock \bibinfo{journal}{\emph{ACM Trans. Graph.}} \bibinfo{volume}{43}, \bibinfo{number}{4}, Article \bibinfo{articleno}{99} (\bibinfo{date}{July} \bibinfo{year}{2024}), \bibinfo{numpages}{18}~pages.
\newblock
\showISSN{0730-0301}
\urldef\tempurl%
\url{https://doi.org/10.1145/3658201}
\showDOI{\tempurl}


\bibitem[Gu et~al\mbox{.}(2002)]%
        {Gu2002GeometryImages}
\bibfield{author}{\bibinfo{person}{Xianfeng Gu}, \bibinfo{person}{Steven~J. Gortler}, {and} \bibinfo{person}{Hugues Hoppe}.} \bibinfo{year}{2002}\natexlab{}.
\newblock \showarticletitle{Geometry images}.
\newblock \bibinfo{journal}{\emph{ACM Trans. Graph.}} \bibinfo{volume}{21}, \bibinfo{number}{3} (\bibinfo{date}{July} \bibinfo{year}{2002}), \bibinfo{pages}{355–361}.
\newblock
\showISSN{0730-0301}
\urldef\tempurl%
\url{https://doi.org/10.1145/566654.566589}
\showDOI{\tempurl}


\bibitem[Guerrero et~al\mbox{.}(2018)]%
        {Guerrero2018PCPNet}
\bibfield{author}{\bibinfo{person}{Paul Guerrero}, \bibinfo{person}{Yanir Kleiman}, \bibinfo{person}{Maks Ovsjanikov}, {and} \bibinfo{person}{Niloy~J. Mitra}.} \bibinfo{year}{2018}\natexlab{}.
\newblock \showarticletitle{PCPN<scp>et</scp> Learning Local Shape Properties from Raw Point Clouds}.
\newblock \bibinfo{journal}{\emph{Computer Graphics Forum}} \bibinfo{volume}{37}, \bibinfo{number}{2} (\bibinfo{date}{May} \bibinfo{year}{2018}), \bibinfo{pages}{75–85}.
\newblock
\showISSN{1467-8659}
\urldef\tempurl%
\url{https://doi.org/10.1111/cgf.13343}
\showDOI{\tempurl}


\bibitem[He et~al\mbox{.}(2015)]%
        {Kaiming2015}
\bibfield{author}{\bibinfo{person}{Kaiming He}, \bibinfo{person}{Xiangyu Zhang}, \bibinfo{person}{Shaoqing Ren}, {and} \bibinfo{person}{Jian Sun}.} \bibinfo{year}{2015}\natexlab{}.
\newblock \showarticletitle{Delving Deep into Rectifiers: Surpassing Human-Level Performance on ImageNet Classification}. In \bibinfo{booktitle}{\emph{2015 IEEE International Conference on Computer Vision (ICCV)}}. \bibinfo{pages}{1026--1034}.
\newblock
\urldef\tempurl%
\url{https://doi.org/10.1109/ICCV.2015.123}
\showDOI{\tempurl}


\bibitem[Hormann and Greiner(2000)]%
        {hormann2000mips}
\bibfield{author}{\bibinfo{person}{Kai Hormann} {and} \bibinfo{person}{G{\"u}nther Greiner}.} \bibinfo{year}{2000}\natexlab{}.
\newblock \showarticletitle{MIPS: An efficient global parametrization method}.
\newblock \bibinfo{journal}{\emph{Curve and Surface Design: Saint-Malo 1999}} (\bibinfo{year}{2000}), \bibinfo{pages}{153--162}.
\newblock


\bibitem[Huang et~al\mbox{.}(2024)]%
        {huang2024surface}
\bibfield{author}{\bibinfo{person}{Zhangjin Huang}, \bibinfo{person}{Yuxin Wen}, \bibinfo{person}{Zihao Wang}, \bibinfo{person}{Jinjuan Ren}, {and} \bibinfo{person}{Kui Jia}.} \bibinfo{year}{2024}\natexlab{}.
\newblock \showarticletitle{Surface reconstruction from point clouds: A survey and a benchmark}.
\newblock \bibinfo{journal}{\emph{IEEE Transactions on Pattern Analysis and Machine Intelligence}} (\bibinfo{year}{2024}).
\newblock


\bibitem[Hwang and Sung(2024)]%
        {Hwang2024ODC}
\bibfield{author}{\bibinfo{person}{Jisung Hwang} {and} \bibinfo{person}{Minhyuk Sung}.} \bibinfo{year}{2024}\natexlab{}.
\newblock \showarticletitle{Occupancy-Based Dual Contouring}. In \bibinfo{booktitle}{\emph{SIGGRAPH Asia 2024 Conference Papers}} \emph{(\bibinfo{series}{SA '24})}. \bibinfo{publisher}{Association for Computing Machinery}, \bibinfo{address}{New York, NY, USA}, Article \bibinfo{articleno}{129}, \bibinfo{numpages}{11}~pages.
\newblock
\showISBNx{9798400711312}
\urldef\tempurl%
\url{https://doi.org/10.1145/3680528.3687581}
\showDOI{\tempurl}


\bibitem[Jacobson et~al\mbox{.}(2013)]%
        {Jacobson2013RobustInside}
\bibfield{author}{\bibinfo{person}{Alec Jacobson}, \bibinfo{person}{Ladislav Kavan}, {and} \bibinfo{person}{Olga Sorkine-Hornung}.} \bibinfo{year}{2013}\natexlab{}.
\newblock \showarticletitle{Robust inside-outside segmentation using generalized winding numbers}.
\newblock \bibinfo{journal}{\emph{ACM Trans. Graph.}} \bibinfo{volume}{32}, \bibinfo{number}{4}, Article \bibinfo{articleno}{33} (\bibinfo{date}{July} \bibinfo{year}{2013}), \bibinfo{numpages}{12}~pages.
\newblock
\showISSN{0730-0301}
\urldef\tempurl%
\url{https://doi.org/10.1145/2461912.2461916}
\showDOI{\tempurl}


\bibitem[Jacobson et~al\mbox{.}(2018)]%
        {libigl}
\bibfield{author}{\bibinfo{person}{Alec Jacobson}, \bibinfo{person}{Daniele Panozzo}, {et~al\mbox{.}}} \bibinfo{year}{2018}\natexlab{}.
\newblock \bibinfo{title}{{libigl}: A simple {C++} geometry processing library}.
\newblock
\newblock
\newblock
\shownote{https://libigl.github.io/}.


\bibitem[Jakob et~al\mbox{.}(2016)]%
        {pybind11}
\bibfield{author}{\bibinfo{person}{Wenzel Jakob}, \bibinfo{person}{Jason Rhinelander}, {and} \bibinfo{person}{Dean Moldovan}.} \bibinfo{year}{2016}\natexlab{}.
\newblock \bibinfo{title}{pybind11 — Seamless operability between C++11 and Python}.
\newblock
\newblock
\newblock
\shownote{https://github.com/pybind/pybind11}.


\bibitem[Ju et~al\mbox{.}(2002)]%
        {Ju2002DualContouring}
\bibfield{author}{\bibinfo{person}{Tao Ju}, \bibinfo{person}{Frank Losasso}, \bibinfo{person}{Scott Schaefer}, {and} \bibinfo{person}{Joe Warren}.} \bibinfo{year}{2002}\natexlab{}.
\newblock \showarticletitle{Dual contouring of hermite data}.
\newblock \bibinfo{journal}{\emph{ACM Trans. Graph.}} \bibinfo{volume}{21}, \bibinfo{number}{3} (\bibinfo{date}{July} \bibinfo{year}{2002}), \bibinfo{pages}{339–346}.
\newblock
\showISSN{0730-0301}
\urldef\tempurl%
\url{https://doi.org/10.1145/566654.566586}
\showDOI{\tempurl}


\bibitem[Kazhdan et~al\mbox{.}(2006)]%
        {kazhdan2006poisson}
\bibfield{author}{\bibinfo{person}{Michael Kazhdan}, \bibinfo{person}{Matthew Bolitho}, {and} \bibinfo{person}{Hugues Hoppe}.} \bibinfo{year}{2006}\natexlab{}.
\newblock \showarticletitle{Poisson surface reconstruction}. In \bibinfo{booktitle}{\emph{Proceedings of the fourth Eurographics symposium on Geometry processing}}, Vol.~\bibinfo{volume}{7}.
\newblock


\bibitem[Kerbl et~al\mbox{.}(2023)]%
        {Kerbl20233DGaussianSplat}
\bibfield{author}{\bibinfo{person}{Bernhard Kerbl}, \bibinfo{person}{Georgios Kopanas}, \bibinfo{person}{Thomas Leimkuehler}, {and} \bibinfo{person}{George Drettakis}.} \bibinfo{year}{2023}\natexlab{}.
\newblock \showarticletitle{3D Gaussian Splatting for Real-Time Radiance Field Rendering}.
\newblock \bibinfo{journal}{\emph{ACM Trans. Graph.}} \bibinfo{volume}{42}, \bibinfo{number}{4}, Article \bibinfo{articleno}{139} (\bibinfo{date}{July} \bibinfo{year}{2023}), \bibinfo{numpages}{14}~pages.
\newblock
\showISSN{0730-0301}
\urldef\tempurl%
\url{https://doi.org/10.1145/3592433}
\showDOI{\tempurl}


\bibitem[King et~al\mbox{.}(2024)]%
        {King2024ClosestPointMethod}
\bibfield{author}{\bibinfo{person}{Nathan King}, \bibinfo{person}{Haozhe Su}, \bibinfo{person}{Mridul Aanjaneya}, \bibinfo{person}{Steven Ruuth}, {and} \bibinfo{person}{Christopher Batty}.} \bibinfo{year}{2024}\natexlab{}.
\newblock \showarticletitle{A Closest Point Method for PDEs on Manifolds with Interior Boundary Conditions for Geometry Processing}.
\newblock \bibinfo{journal}{\emph{ACM Trans. Graph.}} \bibinfo{volume}{43}, \bibinfo{number}{5}, Article \bibinfo{articleno}{159} (\bibinfo{date}{Aug.} \bibinfo{year}{2024}), \bibinfo{numpages}{26}~pages.
\newblock
\showISSN{0730-0301}
\urldef\tempurl%
\url{https://doi.org/10.1145/3673652}
\showDOI{\tempurl}


\bibitem[Kingma and Ba(2017)]%
        {kingma2017adam}
\bibfield{author}{\bibinfo{person}{Diederik~P. Kingma} {and} \bibinfo{person}{Jimmy Ba}.} \bibinfo{year}{2017}\natexlab{}.
\newblock \bibinfo{title}{Adam: A Method for Stochastic Optimization}.
\newblock
\newblock
\showeprint[arxiv]{1412.6980}~[cs.LG]


\bibitem[Kn\"{o}ppel et~al\mbox{.}(2013)]%
        {Knoppel2013GloballyOptimal}
\bibfield{author}{\bibinfo{person}{Felix Kn\"{o}ppel}, \bibinfo{person}{Keenan Crane}, \bibinfo{person}{Ulrich Pinkall}, {and} \bibinfo{person}{Peter Schr\"{o}der}.} \bibinfo{year}{2013}\natexlab{}.
\newblock \showarticletitle{Globally optimal direction fields}.
\newblock \bibinfo{journal}{\emph{ACM Trans. Graph.}} \bibinfo{volume}{32}, \bibinfo{number}{4}, Article \bibinfo{articleno}{59} (\bibinfo{date}{July} \bibinfo{year}{2013}), \bibinfo{numpages}{10}~pages.
\newblock
\showISSN{0730-0301}
\urldef\tempurl%
\url{https://doi.org/10.1145/2461912.2462005}
\showDOI{\tempurl}


\bibitem[Kn\"{o}ppel et~al\mbox{.}(2015)]%
        {Knoppel2015StripePatterns}
\bibfield{author}{\bibinfo{person}{Felix Kn\"{o}ppel}, \bibinfo{person}{Keenan Crane}, \bibinfo{person}{Ulrich Pinkall}, {and} \bibinfo{person}{Peter Schr\"{o}der}.} \bibinfo{year}{2015}\natexlab{}.
\newblock \showarticletitle{Stripe patterns on surfaces}.
\newblock \bibinfo{journal}{\emph{ACM Trans. Graph.}} \bibinfo{volume}{34}, \bibinfo{number}{4}, Article \bibinfo{articleno}{39} (\bibinfo{date}{July} \bibinfo{year}{2015}), \bibinfo{numpages}{11}~pages.
\newblock
\showISSN{0730-0301}
\urldef\tempurl%
\url{https://doi.org/10.1145/2767000}
\showDOI{\tempurl}


\bibitem[Lee et~al\mbox{.}(1998)]%
        {Lee1998maps}
\bibfield{author}{\bibinfo{person}{Aaron W.~F. Lee}, \bibinfo{person}{Wim Sweldens}, \bibinfo{person}{Peter Schr\"{o}der}, \bibinfo{person}{Lawrence Cowsar}, {and} \bibinfo{person}{David Dobkin}.} \bibinfo{year}{1998}\natexlab{}.
\newblock \showarticletitle{MAPS: multiresolution adaptive parameterization of surfaces}. In \bibinfo{booktitle}{\emph{Proceedings of the 25th Annual Conference on Computer Graphics and Interactive Techniques}} \emph{(\bibinfo{series}{SIGGRAPH '98})}. \bibinfo{publisher}{Association for Computing Machinery}, \bibinfo{address}{New York, NY, USA}, \bibinfo{pages}{95–104}.
\newblock
\showISBNx{0897919998}
\urldef\tempurl%
\url{https://doi.org/10.1145/280814.280828}
\showDOI{\tempurl}


\bibitem[Lescoat et~al\mbox{.}(2020)]%
        {lescoat2020spectralmesh}
\bibfield{author}{\bibinfo{person}{Thibault Lescoat}, \bibinfo{person}{Hsueh-Ti~Derek Liu}, \bibinfo{person}{Jean-Marc Thiery}, \bibinfo{person}{Alec Jacobson}, \bibinfo{person}{Tamy Boubekeur}, {and} \bibinfo{person}{Maks Ovsjanikov}.} \bibinfo{year}{2020}\natexlab{}.
\newblock \showarticletitle{Spectral Mesh Simplification}.
\newblock \bibinfo{journal}{\emph{Computer Graphics Forum}} \bibinfo{volume}{39}, \bibinfo{number}{2} (\bibinfo{year}{2020}), \bibinfo{pages}{315--324}.
\newblock
\urldef\tempurl%
\url{https://doi.org/10.1111/cgf.13932}
\showDOI{\tempurl}
\showeprint{https://onlinelibrary.wiley.com/doi/pdf/10.1111/cgf.13932}


\bibitem[Li et~al\mbox{.}(2022)]%
        {li2022compressingvolumetricradiancefields}
\bibfield{author}{\bibinfo{person}{Lingzhi Li}, \bibinfo{person}{Zhen Shen}, \bibinfo{person}{Zhongshu Wang}, \bibinfo{person}{Li Shen}, {and} \bibinfo{person}{Liefeng Bo}.} \bibinfo{year}{2022}\natexlab{}.
\newblock \bibinfo{title}{Compressing Volumetric Radiance Fields to 1 MB}.
\newblock
\newblock
\showeprint[arxiv]{2211.16386}~[cs.CV]
\urldef\tempurl%
\url{https://arxiv.org/abs/2211.16386}
\showURL{%
\tempurl}


\bibitem[Li et~al\mbox{.}(2023)]%
        {li2023closest}
\bibfield{author}{\bibinfo{person}{Mica Li}, \bibinfo{person}{Michael Owens}, \bibinfo{person}{Juheng Wu}, \bibinfo{person}{Grace Yang}, {and} \bibinfo{person}{Albert Chern}.} \bibinfo{year}{2023}\natexlab{}.
\newblock \showarticletitle{Closest Point Exterior Calculus}.
\newblock In \bibinfo{booktitle}{\emph{SIGGRAPH Asia 2023 Posters}}. \bibinfo{pages}{1--2}.
\newblock


\bibitem[Lipman et~al\mbox{.}(2024)]%
        {lipman2024flowmatchingguidecode}
\bibfield{author}{\bibinfo{person}{Yaron Lipman}, \bibinfo{person}{Marton Havasi}, \bibinfo{person}{Peter Holderrieth}, \bibinfo{person}{Neta Shaul}, \bibinfo{person}{Matt Le}, \bibinfo{person}{Brian Karrer}, \bibinfo{person}{Ricky T.~Q. Chen}, \bibinfo{person}{David Lopez-Paz}, \bibinfo{person}{Heli Ben-Hamu}, {and} \bibinfo{person}{Itai Gat}.} \bibinfo{year}{2024}\natexlab{}.
\newblock \bibinfo{title}{Flow Matching Guide and Code}.
\newblock
\newblock
\showeprint[arxiv]{2412.06264}~[cs.LG]
\urldef\tempurl%
\url{https://arxiv.org/abs/2412.06264}
\showURL{%
\tempurl}


\bibitem[Liu et~al\mbox{.}(2020)]%
        {Liu2020NeuralSubdivision}
\bibfield{author}{\bibinfo{person}{Hsueh-Ti~Derek Liu}, \bibinfo{person}{Vladimir~G. Kim}, \bibinfo{person}{Siddhartha Chaudhuri}, \bibinfo{person}{Noam Aigerman}, {and} \bibinfo{person}{Alec Jacobson}.} \bibinfo{year}{2020}\natexlab{}.
\newblock \showarticletitle{Neural subdivision}.
\newblock \bibinfo{journal}{\emph{ACM Trans. Graph.}} \bibinfo{volume}{39}, \bibinfo{number}{4}, Article \bibinfo{articleno}{124} (\bibinfo{date}{Aug.} \bibinfo{year}{2020}), \bibinfo{numpages}{16}~pages.
\newblock
\showISSN{0730-0301}
\urldef\tempurl%
\url{https://doi.org/10.1145/3386569.3392418}
\showDOI{\tempurl}


\bibitem[Liu et~al\mbox{.}(2021)]%
        {Liu2021SurfaceMultigrid}
\bibfield{author}{\bibinfo{person}{Hsueh-Ti~Derek Liu}, \bibinfo{person}{Jiayi~Eris Zhang}, \bibinfo{person}{Mirela Ben-Chen}, {and} \bibinfo{person}{Alec Jacobson}.} \bibinfo{year}{2021}\natexlab{}.
\newblock \showarticletitle{Surface multigrid via intrinsic prolongation}.
\newblock \bibinfo{journal}{\emph{ACM Trans. Graph.}} \bibinfo{volume}{40}, \bibinfo{number}{4}, Article \bibinfo{articleno}{80} (\bibinfo{date}{July} \bibinfo{year}{2021}), \bibinfo{numpages}{13}~pages.
\newblock
\showISSN{0730-0301}
\urldef\tempurl%
\url{https://doi.org/10.1145/3450626.3459768}
\showDOI{\tempurl}


\bibitem[Lopes and Brodlie(2003)]%
        {Lopes2003ImprovingMC}
\bibfield{author}{\bibinfo{person}{Adriano Lopes} {and} \bibinfo{person}{Ken Brodlie}.} \bibinfo{year}{2003}\natexlab{}.
\newblock \showarticletitle{Improving the Robustness and Accuracy of the Marching Cubes Algorithm for Isosurfacing}.
\newblock \bibinfo{journal}{\emph{IEEE Transactions on Visualization and Computer Graphics}} \bibinfo{volume}{9}, \bibinfo{number}{1} (\bibinfo{date}{Jan.} \bibinfo{year}{2003}), \bibinfo{pages}{16–29}.
\newblock
\showISSN{1077-2626}
\urldef\tempurl%
\url{https://doi.org/10.1109/TVCG.2003.1175094}
\showDOI{\tempurl}


\bibitem[Lorensen and Cline(1987)]%
        {Lorensen1987MarchingCubes}
\bibfield{author}{\bibinfo{person}{William~E. Lorensen} {and} \bibinfo{person}{Harvey~E. Cline}.} \bibinfo{year}{1987}\natexlab{}.
\newblock \showarticletitle{Marching cubes: A high resolution 3D surface construction algorithm}.
\newblock \bibinfo{journal}{\emph{SIGGRAPH Comput. Graph.}} \bibinfo{volume}{21}, \bibinfo{number}{4} (\bibinfo{date}{Aug.} \bibinfo{year}{1987}), \bibinfo{pages}{163–169}.
\newblock
\showISSN{0097-8930}
\urldef\tempurl%
\url{https://doi.org/10.1145/37402.37422}
\showDOI{\tempurl}


\bibitem[Macklin(2022)]%
        {warp2022}
\bibfield{author}{\bibinfo{person}{Miles Macklin}.} \bibinfo{year}{2022}\natexlab{}.
\newblock \bibinfo{title}{Warp: A High-performance Python Framework for GPU Simulation and Graphics}.
\newblock \bibinfo{howpublished}{\url{https://github.com/nvidia/warp}}.
\newblock
\newblock
\shownote{NVIDIA GPU Technology Conference (GTC)}.


\bibitem[Marschner et~al\mbox{.}(2023)]%
        {Marschner2023ConstructiveSolid}
\bibfield{author}{\bibinfo{person}{Zo\"{e} Marschner}, \bibinfo{person}{Silvia Sell\'{a}n}, \bibinfo{person}{Hsueh-Ti~Derek Liu}, {and} \bibinfo{person}{Alec Jacobson}.} \bibinfo{year}{2023}\natexlab{}.
\newblock \showarticletitle{Constructive Solid Geometry on Neural Signed Distance Fields}. In \bibinfo{booktitle}{\emph{SIGGRAPH Asia 2023 Conference Papers}} (Sydney, NSW, Australia) \emph{(\bibinfo{series}{SA '23})}. \bibinfo{publisher}{Association for Computing Machinery}, \bibinfo{address}{New York, NY, USA}, Article \bibinfo{articleno}{121}, \bibinfo{numpages}{12}~pages.
\newblock
\showISBNx{9798400703157}
\urldef\tempurl%
\url{https://doi.org/10.1145/3610548.3618170}
\showDOI{\tempurl}


\bibitem[Mildenhall et~al\mbox{.}(2020)]%
        {mildenhall2020nerf}
\bibfield{author}{\bibinfo{person}{Ben Mildenhall}, \bibinfo{person}{Pratul~P. Srinivasan}, \bibinfo{person}{Matthew Tancik}, \bibinfo{person}{Jonathan~T. Barron}, \bibinfo{person}{Ravi Ramamoorthi}, {and} \bibinfo{person}{Ren Ng}.} \bibinfo{year}{2020}\natexlab{}.
\newblock \showarticletitle{NeRF: Representing Scenes as Neural Radiance Fields for View Synthesis}. In \bibinfo{booktitle}{\emph{ECCV}}.
\newblock


\bibitem[Mitchell et~al\mbox{.}(1987)]%
        {mitchell1987mmp}
\bibfield{author}{\bibinfo{person}{Joseph S.~B. Mitchell}, \bibinfo{person}{David~M. Mount}, {and} \bibinfo{person}{Christos~H. Papadimitriou}.} \bibinfo{year}{1987}\natexlab{}.
\newblock \showarticletitle{The Discrete Geodesic Problem}.
\newblock \bibinfo{journal}{\emph{SIAM J. Comput.}} \bibinfo{volume}{16}, \bibinfo{number}{4} (\bibinfo{year}{1987}), \bibinfo{pages}{647--668}.
\newblock
\urldef\tempurl%
\url{https://doi.org/10.1137/0216045}
\showDOI{\tempurl}
\showeprint{https://doi.org/10.1137/0216045}


\bibitem[Morreale et~al\mbox{.}(2021)]%
        {Morreale2021NeuralSurfaceMaps}
\bibfield{author}{\bibinfo{person}{Luca Morreale}, \bibinfo{person}{Noam Aigerman}, \bibinfo{person}{Vladimir Kim}, {and} \bibinfo{person}{Niloy~J. Mitra}.} \bibinfo{year}{2021}\natexlab{}.
\newblock \showarticletitle{Neural Surface Maps}. In \bibinfo{booktitle}{\emph{2021 IEEE/CVF Conference on Computer Vision and Pattern Recognition (CVPR)}}. \bibinfo{pages}{4637--4646}.
\newblock
\urldef\tempurl%
\url{https://doi.org/10.1109/CVPR46437.2021.00461}
\showDOI{\tempurl}


\bibitem[Müller et~al\mbox{.}(2022)]%
        {Muller2022InstantNGP}
\bibfield{author}{\bibinfo{person}{Thomas Müller}, \bibinfo{person}{Alex Evans}, \bibinfo{person}{Christoph Schied}, {and} \bibinfo{person}{Alexander Keller}.} \bibinfo{year}{2022}\natexlab{}.
\newblock \showarticletitle{Instant neural graphics primitives with a multiresolution hash encoding}.
\newblock \bibinfo{journal}{\emph{ACM Transactions on Graphics}} \bibinfo{volume}{41}, \bibinfo{number}{4} (\bibinfo{date}{July} \bibinfo{year}{2022}), \bibinfo{pages}{1–15}.
\newblock
\showISSN{1557-7368}
\urldef\tempurl%
\url{https://doi.org/10.1145/3528223.3530127}
\showDOI{\tempurl}


\bibitem[Novello et~al\mbox{.}(2022)]%
        {Novello2022ExploringDifferentialGeometry}
\bibfield{author}{\bibinfo{person}{Tiago Novello}, \bibinfo{person}{Guilherme Schardong}, \bibinfo{person}{Luiz Schirmer}, \bibinfo{person}{Vin\'{\i}cius da Silva}, \bibinfo{person}{H\'{e}lio Lopes}, {and} \bibinfo{person}{Luiz Velho}.} \bibinfo{year}{2022}\natexlab{}.
\newblock \showarticletitle{Exploring differential geometry in neural implicits}.
\newblock \bibinfo{journal}{\emph{Comput. Graph.}} \bibinfo{volume}{108}, \bibinfo{number}{C} (\bibinfo{date}{Nov.} \bibinfo{year}{2022}), \bibinfo{pages}{49–60}.
\newblock
\showISSN{0097-8493}
\urldef\tempurl%
\url{https://doi.org/10.1016/j.cag.2022.09.003}
\showDOI{\tempurl}


\bibitem[Otaduy and Lin(2003)]%
        {otaduy2003sensationpreserving}
\bibfield{author}{\bibinfo{person}{Miguel~A. Otaduy} {and} \bibinfo{person}{Ming~C. Lin}.} \bibinfo{year}{2003}\natexlab{}.
\newblock \showarticletitle{Sensation preserving simplification for haptic rendering}.
\newblock \bibinfo{journal}{\emph{ACM Trans. Graph.}} \bibinfo{volume}{22}, \bibinfo{number}{3} (\bibinfo{date}{July} \bibinfo{year}{2003}), \bibinfo{pages}{543–553}.
\newblock
\showISSN{0730-0301}
\urldef\tempurl%
\url{https://doi.org/10.1145/882262.882305}
\showDOI{\tempurl}


\bibitem[Pang et~al\mbox{.}(2024)]%
        {Pang2024NeuralLaplacianOperator}
\bibfield{author}{\bibinfo{person}{Bo Pang}, \bibinfo{person}{Zhongtian Zheng}, \bibinfo{person}{Yilong Li}, \bibinfo{person}{Guoping Wang}, {and} \bibinfo{person}{Peng-Shuai Wang}.} \bibinfo{year}{2024}\natexlab{}.
\newblock \showarticletitle{Neural Laplacian Operator for 3D Point Clouds}.
\newblock \bibinfo{journal}{\emph{ACM Trans. Graph.}} \bibinfo{volume}{43}, \bibinfo{number}{6}, Article \bibinfo{articleno}{239} (\bibinfo{date}{Nov.} \bibinfo{year}{2024}), \bibinfo{numpages}{14}~pages.
\newblock
\showISSN{0730-0301}
\urldef\tempurl%
\url{https://doi.org/10.1145/3687901}
\showDOI{\tempurl}


\bibitem[Park et~al\mbox{.}(2019)]%
        {park2019deepsdflearningcontinuoussigned}
\bibfield{author}{\bibinfo{person}{Jeong~Joon Park}, \bibinfo{person}{Peter Florence}, \bibinfo{person}{Julian Straub}, \bibinfo{person}{Richard Newcombe}, {and} \bibinfo{person}{Steven Lovegrove}.} \bibinfo{year}{2019}\natexlab{}.
\newblock \bibinfo{title}{DeepSDF: Learning Continuous Signed Distance Functions for Shape Representation}.
\newblock
\newblock
\showeprint[arxiv]{1901.05103}~[cs.CV]
\urldef\tempurl%
\url{https://arxiv.org/abs/1901.05103}
\showURL{%
\tempurl}


\bibitem[Paszke et~al\mbox{.}(2017)]%
        {pytorch}
\bibfield{author}{\bibinfo{person}{Adam Paszke}, \bibinfo{person}{Sam Gross}, \bibinfo{person}{Soumith Chintala}, \bibinfo{person}{Gregory Chanan}, \bibinfo{person}{Edward Yang}, \bibinfo{person}{Zachary DeVito}, \bibinfo{person}{Zeming Lin}, \bibinfo{person}{Alban Desmaison}, \bibinfo{person}{Luca Antiga}, {and} \bibinfo{person}{Adam Lerer}.} \bibinfo{year}{2017}\natexlab{}.
\newblock \showarticletitle{Automatic differentiation in PyTorch}.
\newblock  (\bibinfo{year}{2017}).
\newblock


\bibitem[Pentapati et~al\mbox{.}(2025)]%
        {pentapati2025meshcompressionquantizedneural}
\bibfield{author}{\bibinfo{person}{Sai~Karthikey Pentapati}, \bibinfo{person}{Gregoire Phillips}, {and} \bibinfo{person}{Alan~C. Bovik}.} \bibinfo{year}{2025}\natexlab{}.
\newblock \bibinfo{title}{Mesh Compression with Quantized Neural Displacement Fields}.
\newblock
\newblock
\showeprint[arxiv]{2504.01027}~[cs.GR]
\urldef\tempurl%
\url{https://arxiv.org/abs/2504.01027}
\showURL{%
\tempurl}


\bibitem[Pistilli et~al\mbox{.}(2020)]%
        {Pistilli2020PointCloudNormal}
\bibfield{author}{\bibinfo{person}{Francesca Pistilli}, \bibinfo{person}{Giulia Fracastoro}, \bibinfo{person}{Diego Valsesia}, {and} \bibinfo{person}{Enrico Magli}.} \bibinfo{year}{2020}\natexlab{}.
\newblock \showarticletitle{Point Cloud Normal Estimation with Graph-Convolutional Neural Networks}. In \bibinfo{booktitle}{\emph{2020 IEEE International Conference on Multimedia \& Expo Workshops (ICMEW)}}. \bibinfo{pages}{1--6}.
\newblock
\urldef\tempurl%
\url{https://doi.org/10.1109/ICMEW46912.2020.9105972}
\showDOI{\tempurl}


\bibitem[Rossignac(2001)]%
        {rossignac20013dcompression}
\bibfield{author}{\bibinfo{person}{Jarek Rossignac}.} \bibinfo{year}{2001}\natexlab{}.
\newblock \showarticletitle{3D Compression Made Simple: Edgebreaker with Zip\&Wrap on a Corner-Table}. In \bibinfo{booktitle}{\emph{Proceedings of the International Conference on Shape Modeling \& Applications}} \emph{(\bibinfo{series}{SMI '01})}. \bibinfo{publisher}{IEEE Computer Society}, \bibinfo{address}{USA}, \bibinfo{pages}{278}.
\newblock
\showISBNx{0769508537}


\bibitem[Sawhney and Crane(2020)]%
        {Sawhney2020MonteCarloGeometry}
\bibfield{author}{\bibinfo{person}{Rohan Sawhney} {and} \bibinfo{person}{Keenan Crane}.} \bibinfo{year}{2020}\natexlab{}.
\newblock \showarticletitle{Monte Carlo geometry processing: a grid-free approach to PDE-based methods on volumetric domains}.
\newblock \bibinfo{journal}{\emph{ACM Trans. Graph.}} \bibinfo{volume}{39}, \bibinfo{number}{4}, Article \bibinfo{articleno}{123} (\bibinfo{date}{Aug.} \bibinfo{year}{2020}), \bibinfo{numpages}{18}~pages.
\newblock
\showISSN{0730-0301}
\urldef\tempurl%
\url{https://doi.org/10.1145/3386569.3392374}
\showDOI{\tempurl}


\bibitem[Sawhney et~al\mbox{.}(2022)]%
        {Sawhney2022Grid-freeMonteCarlo}
\bibfield{author}{\bibinfo{person}{Rohan Sawhney}, \bibinfo{person}{Dario Seyb}, \bibinfo{person}{Wojciech Jarosz}, {and} \bibinfo{person}{Keenan Crane}.} \bibinfo{year}{2022}\natexlab{}.
\newblock \showarticletitle{Grid-free Monte Carlo for PDEs with spatially varying coefficients}.
\newblock \bibinfo{journal}{\emph{ACM Trans. Graph.}} \bibinfo{volume}{41}, \bibinfo{number}{4}, Article \bibinfo{articleno}{53} (\bibinfo{date}{July} \bibinfo{year}{2022}), \bibinfo{numpages}{17}~pages.
\newblock
\showISSN{0730-0301}
\urldef\tempurl%
\url{https://doi.org/10.1145/3528223.3530134}
\showDOI{\tempurl}


\bibitem[Schaefer et~al\mbox{.}(2007)]%
        {Schaefer2007ManifoldDC}
\bibfield{author}{\bibinfo{person}{Scott Schaefer}, \bibinfo{person}{Tao Ju}, {and} \bibinfo{person}{Joe Warren}.} \bibinfo{year}{2007}\natexlab{}.
\newblock \showarticletitle{Manifold Dual Contouring}.
\newblock \bibinfo{journal}{\emph{IEEE Transactions on Visualization and Computer Graphics}} \bibinfo{volume}{13}, \bibinfo{number}{3} (\bibinfo{date}{May} \bibinfo{year}{2007}), \bibinfo{pages}{610–619}.
\newblock
\showISSN{1077-2626}
\urldef\tempurl%
\url{https://doi.org/10.1109/TVCG.2007.1012}
\showDOI{\tempurl}


\bibitem[Schaefer and Warren(2004)]%
        {Schaefer2004DualMarchingCubes}
\bibfield{author}{\bibinfo{person}{Scott Schaefer} {and} \bibinfo{person}{Joe Warren}.} \bibinfo{year}{2004}\natexlab{}.
\newblock \showarticletitle{Dual Marching Cubes: Primal Contouring of Dual Grids}. In \bibinfo{booktitle}{\emph{Proceedings of the Computer Graphics and Applications, 12th Pacific Conference}} \emph{(\bibinfo{series}{PG '04})}. \bibinfo{publisher}{IEEE Computer Society}, \bibinfo{address}{USA}, \bibinfo{pages}{70–76}.
\newblock
\showISBNx{0769522343}


\bibitem[Schmidt et~al\mbox{.}(2023)]%
        {Schmidt2023SurfaceMaps}
\bibfield{author}{\bibinfo{person}{P. Schmidt}, \bibinfo{person}{D. Pieper}, {and} \bibinfo{person}{L. Kobbelt}.} \bibinfo{year}{2023}\natexlab{}.
\newblock \showarticletitle{Surface Maps via Adaptive Triangulations}.
\newblock \bibinfo{journal}{\emph{Computer Graphics Forum}} \bibinfo{volume}{42}, \bibinfo{number}{2} (\bibinfo{year}{2023}), \bibinfo{pages}{103--117}.
\newblock
\urldef\tempurl%
\url{https://doi.org/10.1111/cgf.14747}
\showDOI{\tempurl}
\showeprint{https://onlinelibrary.wiley.com/doi/pdf/10.1111/cgf.14747}


\bibitem[Sell\'{a}n et~al\mbox{.}(2023a)]%
        {Sellan2023BreakingGood}
\bibfield{author}{\bibinfo{person}{Silvia Sell\'{a}n}, \bibinfo{person}{Jack Luong}, \bibinfo{person}{Leticia Mattos Da~Silva}, \bibinfo{person}{Aravind Ramakrishnan}, \bibinfo{person}{Yuchuan Yang}, {and} \bibinfo{person}{Alec Jacobson}.} \bibinfo{year}{2023}\natexlab{a}.
\newblock \showarticletitle{Breaking Good: Fracture Modes for Realtime Destruction}.
\newblock \bibinfo{journal}{\emph{ACM Trans. Graph.}} \bibinfo{volume}{42}, \bibinfo{number}{1}, Article \bibinfo{articleno}{10} (\bibinfo{date}{March} \bibinfo{year}{2023}), \bibinfo{numpages}{12}~pages.
\newblock
\showISSN{0730-0301}
\urldef\tempurl%
\url{https://doi.org/10.1145/3549540}
\showDOI{\tempurl}


\bibitem[Sell\'{a}n et~al\mbox{.}(2023b)]%
        {gpytoolbox}
\bibfield{author}{\bibinfo{person}{Silvia Sell\'{a}n}, \bibinfo{person}{Oded Stein}, {et~al\mbox{.}}} \bibinfo{year}{2023}\natexlab{b}.
\newblock \bibinfo{title}{{gptyoolbox}: A Python Geometry Processing Toolbox}.
\newblock
\newblock
\newblock
\shownote{https://gpytoolbox.org/}.


\bibitem[Seyb et~al\mbox{.}(2019)]%
        {Seyb2019NonlinearSphere}
\bibfield{author}{\bibinfo{person}{Dario Seyb}, \bibinfo{person}{Alec Jacobson}, \bibinfo{person}{Derek Nowrouzezahrai}, {and} \bibinfo{person}{Wojciech Jarosz}.} \bibinfo{year}{2019}\natexlab{}.
\newblock \showarticletitle{Non-linear sphere tracing for rendering deformed signed distance fields}.
\newblock \bibinfo{journal}{\emph{ACM Trans. Graph.}} \bibinfo{volume}{38}, \bibinfo{number}{6}, Article \bibinfo{articleno}{229} (\bibinfo{date}{Nov.} \bibinfo{year}{2019}), \bibinfo{numpages}{12}~pages.
\newblock
\showISSN{0730-0301}
\urldef\tempurl%
\url{https://doi.org/10.1145/3355089.3356502}
\showDOI{\tempurl}


\bibitem[Sharp and Crane(2020)]%
        {Sharp2020LaplacianNonmanifold}
\bibfield{author}{\bibinfo{person}{Nicholas Sharp} {and} \bibinfo{person}{Keenan Crane}.} \bibinfo{year}{2020}\natexlab{}.
\newblock \showarticletitle{A Laplacian for Nonmanifold Triangle Meshes}.
\newblock \bibinfo{journal}{\emph{Computer Graphics Forum}} \bibinfo{volume}{39}, \bibinfo{number}{5} (\bibinfo{year}{2020}), \bibinfo{pages}{69--80}.
\newblock
\urldef\tempurl%
\url{https://doi.org/10.1111/cgf.14069}
\showDOI{\tempurl}
\showeprint{https://onlinelibrary.wiley.com/doi/pdf/10.1111/cgf.14069}


\bibitem[Sharp et~al\mbox{.}(2019a)]%
        {geometrycentral}
\bibfield{author}{\bibinfo{person}{Nicholas Sharp}, \bibinfo{person}{Keenan Crane}, {et~al\mbox{.}}} \bibinfo{year}{2019}\natexlab{a}.
\newblock \showarticletitle{GeometryCentral: A modern C++ library of data structures and algorithms for geometry processing}.
\newblock \bibinfo{howpublished}{\url{https://geometry-central.net/}}.
\newblock  (\bibinfo{year}{2019}).
\newblock


\bibitem[Sharp et~al\mbox{.}(2019b)]%
        {Sharp2019NavigatingIntrinsic}
\bibfield{author}{\bibinfo{person}{Nicholas Sharp}, \bibinfo{person}{Yousuf Soliman}, {and} \bibinfo{person}{Keenan Crane}.} \bibinfo{year}{2019}\natexlab{b}.
\newblock \showarticletitle{Navigating intrinsic triangulations}.
\newblock \bibinfo{journal}{\emph{ACM Trans. Graph.}} \bibinfo{volume}{38}, \bibinfo{number}{4}, Article \bibinfo{articleno}{55} (\bibinfo{date}{July} \bibinfo{year}{2019}), \bibinfo{numpages}{16}~pages.
\newblock
\showISSN{0730-0301}
\urldef\tempurl%
\url{https://doi.org/10.1145/3306346.3322979}
\showDOI{\tempurl}


\bibitem[Sivaram et~al\mbox{.}(2024)]%
        {Sivaram2024NeuralGeometryFields}
\bibfield{author}{\bibinfo{person}{Venkataram~Edavamadathil Sivaram}, \bibinfo{person}{Tzu-Mao Li}, {and} \bibinfo{person}{Ravi Ramamoorthi}.} \bibinfo{year}{2024}\natexlab{}.
\newblock \showarticletitle{Neural Geometry Fields For Meshes}. In \bibinfo{booktitle}{\emph{ACM SIGGRAPH 2024 Conference Papers}} (Denver, CO, USA) \emph{(\bibinfo{series}{SIGGRAPH '24})}. \bibinfo{publisher}{Association for Computing Machinery}, \bibinfo{address}{New York, NY, USA}, Article \bibinfo{articleno}{29}, \bibinfo{numpages}{11}~pages.
\newblock
\showISBNx{9798400705250}
\urldef\tempurl%
\url{https://doi.org/10.1145/3641519.3657399}
\showDOI{\tempurl}


\bibitem[Sugimoto et~al\mbox{.}(2024)]%
        {Sugimoto2024ProjectedWoS}
\bibfield{author}{\bibinfo{person}{Ryusuke Sugimoto}, \bibinfo{person}{Nathan King}, \bibinfo{person}{Toshiya Hachisuka}, {and} \bibinfo{person}{Christopher Batty}.} \bibinfo{year}{2024}\natexlab{}.
\newblock \showarticletitle{Projected Walk on Spheres: A Monte Carlo Closest Point Method for Surface PDEs}. In \bibinfo{booktitle}{\emph{SIGGRAPH Asia 2024 Conference Papers}} \emph{(\bibinfo{series}{SA '24})}. \bibinfo{publisher}{Association for Computing Machinery}, \bibinfo{address}{New York, NY, USA}, Article \bibinfo{articleno}{8}, \bibinfo{numpages}{10}~pages.
\newblock
\showISBNx{9798400711312}
\urldef\tempurl%
\url{https://doi.org/10.1145/3680528.3687599}
\showDOI{\tempurl}


\bibitem[Sun et~al\mbox{.}(2024)]%
        {sun2024tuttenet}
\bibfield{author}{\bibinfo{person}{Bo Sun}, \bibinfo{person}{Thibault Groueix}, \bibinfo{person}{Chen Song}, \bibinfo{person}{Qixing Huang}, {and} \bibinfo{person}{Noam Aigerman}.} \bibinfo{year}{2024}\natexlab{}.
\newblock \showarticletitle{TutteNet: Injective 3D Deformations by Composition of 2D Mesh Deformations}. In \bibinfo{booktitle}{\emph{The IEEE/CVF Conference on Computer Vision and Pattern Recognition 2024}}.
\newblock


\bibitem[Sundararaman et~al\mbox{.}(2024)]%
        {sundararaman2024selfsuperviseddualcontouring}
\bibfield{author}{\bibinfo{person}{Ramana Sundararaman}, \bibinfo{person}{Roman Klokov}, {and} \bibinfo{person}{Maks Ovsjanikov}.} \bibinfo{year}{2024}\natexlab{}.
\newblock \bibinfo{title}{Self-Supervised Dual Contouring}.
\newblock
\newblock
\showeprint[arxiv]{2405.18131}~[cs.CV]
\urldef\tempurl%
\url{https://arxiv.org/abs/2405.18131}
\showURL{%
\tempurl}


\bibitem[Surazhsky and Gotsman(2003)]%
        {surazhsky2003explicitsurface}
\bibfield{author}{\bibinfo{person}{Vitaly Surazhsky} {and} \bibinfo{person}{Craig Gotsman}.} \bibinfo{year}{2003}\natexlab{}.
\newblock \showarticletitle{Explicit surface remeshing}. In \bibinfo{booktitle}{\emph{Proceedings of the 2003 Eurographics/ACM SIGGRAPH Symposium on Geometry Processing}} (Aachen, Germany) \emph{(\bibinfo{series}{SGP '03})}. \bibinfo{publisher}{Eurographics Association}, \bibinfo{address}{Goslar, DEU}, \bibinfo{pages}{20–30}.
\newblock
\showISBNx{1581136870}


\bibitem[Szymczak et~al\mbox{.}(2001)]%
        {szymczak2001edgebreaker}
\bibfield{author}{\bibinfo{person}{Andrzej Szymczak}, \bibinfo{person}{Davis King}, {and} \bibinfo{person}{Jarek Rossignac}.} \bibinfo{year}{2001}\natexlab{}.
\newblock \showarticletitle{An Edgebreaker-based efficient compression scheme for regular meshes}.
\newblock \bibinfo{journal}{\emph{Computational Geometry}} \bibinfo{volume}{20}, \bibinfo{number}{1} (\bibinfo{year}{2001}), \bibinfo{pages}{53--68}.
\newblock
\showISSN{0925-7721}
\urldef\tempurl%
\url{https://doi.org/10.1016/S0925-7721(01)00035-9}
\showDOI{\tempurl}
\newblock
\shownote{Selected papers from the 12th Annual Canadian Conference on}.


\bibitem[Szymczak et~al\mbox{.}(2002)]%
        {szymczak2002piecewiseregular}
\bibfield{author}{\bibinfo{person}{Andrzej Szymczak}, \bibinfo{person}{Jarek Rossignac}, {and} \bibinfo{person}{Davis King}.} \bibinfo{year}{2002}\natexlab{}.
\newblock \showarticletitle{Piecewise Regular Meshes: Construction and Compression}.
\newblock \bibinfo{journal}{\emph{Graphical Models}} \bibinfo{volume}{64}, \bibinfo{number}{3} (\bibinfo{year}{2002}), \bibinfo{pages}{183--198}.
\newblock
\showISSN{1524-0703}
\urldef\tempurl%
\url{https://doi.org/10.1006/gmod.2002.0577}
\showDOI{\tempurl}


\bibitem[Takikawa et~al\mbox{.}(2023)]%
        {takikawa2023compactngp}
\bibfield{author}{\bibinfo{person}{Towaki Takikawa}, \bibinfo{person}{Thomas M\"{u}ller}, \bibinfo{person}{Merlin Nimier-David}, \bibinfo{person}{Alex Evans}, \bibinfo{person}{Sanja Fidler}, \bibinfo{person}{Alec Jacobson}, {and} \bibinfo{person}{Alexander Keller}.} \bibinfo{year}{2023}\natexlab{}.
\newblock \showarticletitle{Compact Neural Graphics Primitives with Learned Hash Probing}. In \bibinfo{booktitle}{\emph{SIGGRAPH Asia 2023 Conference Papers}} (Sydney, NSW, Australia) \emph{(\bibinfo{series}{SA '23})}. \bibinfo{publisher}{Association for Computing Machinery}, \bibinfo{address}{New York, NY, USA}, Article \bibinfo{articleno}{120}, \bibinfo{numpages}{10}~pages.
\newblock
\showISBNx{9798400703157}
\urldef\tempurl%
\url{https://doi.org/10.1145/3610548.3618167}
\showDOI{\tempurl}


\bibitem[Taubin and Rossignac(1998)]%
        {taubin1998geometriccompression}
\bibfield{author}{\bibinfo{person}{Gabriel Taubin} {and} \bibinfo{person}{Jarek Rossignac}.} \bibinfo{year}{1998}\natexlab{}.
\newblock \showarticletitle{Geometric compression through topological surgery}.
\newblock \bibinfo{journal}{\emph{ACM Trans. Graph.}} \bibinfo{volume}{17}, \bibinfo{number}{2} (\bibinfo{date}{April} \bibinfo{year}{1998}), \bibinfo{pages}{84–115}.
\newblock
\showISSN{0730-0301}
\urldef\tempurl%
\url{https://doi.org/10.1145/274363.274365}
\showDOI{\tempurl}


\bibitem[Touma and Gotsman(1998)]%
        {touma1998trianglemesh}
\bibfield{author}{\bibinfo{person}{Costa Touma} {and} \bibinfo{person}{Craig Gotsman}.} \bibinfo{year}{1998}\natexlab{}.
\newblock \showarticletitle{Triangle Mesh Compression}. In \bibinfo{booktitle}{\emph{Proceedings of the Graphics Interface 1998 Conference, June 18-20, 1998, Vancouver, BC, Canada}}. \bibinfo{pages}{26--34}.
\newblock
\urldef\tempurl%
\url{http://graphicsinterface.org/wp-content/uploads/gi1998-4.pdf}
\showURL{%
\tempurl}


\bibitem[Wang et~al\mbox{.}(2021)]%
        {wang2021neus}
\bibfield{author}{\bibinfo{person}{Peng Wang}, \bibinfo{person}{Lingjie Liu}, \bibinfo{person}{Yuan Liu}, \bibinfo{person}{Christian Theobalt}, \bibinfo{person}{Taku Komura}, {and} \bibinfo{person}{Wenping Wang}.} \bibinfo{year}{2021}\natexlab{}.
\newblock \showarticletitle{Neus: Learning neural implicit surfaces by volume rendering for multi-view reconstruction}.
\newblock \bibinfo{journal}{\emph{arXiv preprint arXiv:2106.10689}} (\bibinfo{year}{2021}).
\newblock


\bibitem[Wardetzky et~al\mbox{.}(2007)]%
        {wardetzky2007nofreelunch}
\bibfield{author}{\bibinfo{person}{Max Wardetzky}, \bibinfo{person}{Saurabh Mathur}, \bibinfo{person}{Felix K\"{a}lberer}, {and} \bibinfo{person}{Eitan Grinspun}.} \bibinfo{year}{2007}\natexlab{}.
\newblock \showarticletitle{Discrete laplace operators: no free lunch}. In \bibinfo{booktitle}{\emph{Proceedings of the Fifth Eurographics Symposium on Geometry Processing}} (Barcelona, Spain) \emph{(\bibinfo{series}{SGP '07})}. \bibinfo{publisher}{Eurographics Association}, \bibinfo{address}{Goslar, DEU}, \bibinfo{pages}{33–37}.
\newblock
\showISBNx{9783905673463}


\bibitem[Williamson and Mitra(2024)]%
        {williamson2024neuralgeometryprocessingspherical}
\bibfield{author}{\bibinfo{person}{Romy Williamson} {and} \bibinfo{person}{Niloy~J. Mitra}.} \bibinfo{year}{2024}\natexlab{}.
\newblock \bibinfo{title}{Neural Geometry Processing via Spherical Neural Surfaces}.
\newblock
\newblock
\showeprint[arxiv]{2407.07755}~[cs.GR]
\urldef\tempurl%
\url{https://arxiv.org/abs/2407.07755}
\showURL{%
\tempurl}


\bibitem[Yang et~al\mbox{.}(2021)]%
        {Yang2021NeuralFields}
\bibfield{author}{\bibinfo{person}{Guandao Yang}, \bibinfo{person}{Serge Belongie}, \bibinfo{person}{Bharath Hariharan}, {and} \bibinfo{person}{Vladlen Koltun}.} \bibinfo{year}{2021}\natexlab{}.
\newblock \showarticletitle{Geometry Processing with Neural Fields}. In \bibinfo{booktitle}{\emph{Advances in Neural Information Processing Systems}}, \bibfield{editor}{\bibinfo{person}{M.~Ranzato}, \bibinfo{person}{A.~Beygelzimer}, \bibinfo{person}{Y.~Dauphin}, \bibinfo{person}{P.S. Liang}, {and} \bibinfo{person}{J.~Wortman Vaughan}} (Eds.), Vol.~\bibinfo{volume}{34}. \bibinfo{publisher}{Curran Associates, Inc.}, \bibinfo{pages}{22483--22497}.
\newblock
\urldef\tempurl%
\url{https://proceedings.neurips.cc/paper_files/paper/2021/file/bd686fd640be98efaae0091fa301e613-Paper.pdf}
\showURL{%
\tempurl}


\bibitem[Yu et~al\mbox{.}(2022)]%
        {Yu2022SDFStudio}
\bibfield{author}{\bibinfo{person}{Zehao Yu}, \bibinfo{person}{Anpei Chen}, \bibinfo{person}{Bozidar Antic}, \bibinfo{person}{Songyou Peng}, \bibinfo{person}{Apratim Bhattacharyya}, \bibinfo{person}{Michael Niemeyer}, \bibinfo{person}{Siyu Tang}, \bibinfo{person}{Torsten Sattler}, {and} \bibinfo{person}{Andreas Geiger}.} \bibinfo{year}{2022}\natexlab{}.
\newblock \bibinfo{title}{SDFStudio: A Unified Framework for Surface Reconstruction}.
\newblock
\newblock
\urldef\tempurl%
\url{https://github.com/autonomousvision/sdfstudio}
\showURL{%
\tempurl}


\bibitem[Zhang et~al\mbox{.}(2024)]%
        {zhang2024progressivedynamics}
\bibfield{author}{\bibinfo{person}{Jiayi~Eris Zhang}, \bibinfo{person}{Doug James}, {and} \bibinfo{person}{Danny~M. Kaufman}.} \bibinfo{year}{2024}\natexlab{}.
\newblock \showarticletitle{Progressive Dynamics for Cloth and Shell Animation}.
\newblock \bibinfo{journal}{\emph{ACM Trans. Graph.}} \bibinfo{volume}{43}, \bibinfo{number}{4}, Article \bibinfo{articleno}{104} (\bibinfo{date}{July} \bibinfo{year}{2024}), \bibinfo{numpages}{18}~pages.
\newblock
\showISSN{0730-0301}
\urldef\tempurl%
\url{https://doi.org/10.1145/3658214}
\showDOI{\tempurl}


\bibitem[Zhang et~al\mbox{.}(2022)]%
        {zhang2022nerfusionfusingradiancefields}
\bibfield{author}{\bibinfo{person}{Xiaoshuai Zhang}, \bibinfo{person}{Sai Bi}, \bibinfo{person}{Kalyan Sunkavalli}, \bibinfo{person}{Hao Su}, {and} \bibinfo{person}{Zexiang Xu}.} \bibinfo{year}{2022}\natexlab{}.
\newblock \bibinfo{title}{NeRFusion: Fusing Radiance Fields for Large-Scale Scene Reconstruction}.
\newblock
\newblock
\showeprint[arxiv]{2203.11283}~[cs.CV]
\urldef\tempurl%
\url{https://arxiv.org/abs/2203.11283}
\showURL{%
\tempurl}


\bibitem[Zhou and Jacobson(2016)]%
        {zhou2016thingi10kdataset100003dprinting}
\bibfield{author}{\bibinfo{person}{Qingnan Zhou} {and} \bibinfo{person}{Alec Jacobson}.} \bibinfo{year}{2016}\natexlab{}.
\newblock \bibinfo{title}{Thingi10K: A Dataset of 10,000 3D-Printing Models}.
\newblock
\newblock
\showeprint[arxiv]{1605.04797}~[cs.GR]
\urldef\tempurl%
\url{https://arxiv.org/abs/1605.04797}
\showURL{%
\tempurl}


\end{thebibliography}
